\documentclass[prb,twocolumn,unsortedaddress,superscriptaddress]{revtex4-1}
\usepackage[T1]{fontenc}
\usepackage[utf8]{inputenc}
\usepackage{graphicx}
\usepackage{amsmath}
\usepackage{amssymb}
\usepackage{color}
\usepackage{comment}
\usepackage{multirow}
\usepackage{hyperref}
\usepackage{subcaption}
\usepackage[left=1.5cm,right=1.5cm,top=2cm,bottom=2cm]{geometry}

\newcommand{\ph}[1]{\phantom{#1}}
\newcommand*\diff{\mathop{}\!\mathrm{d}}
\def\ket #1 {\vert {#1} \rangle}
\def\bra #1 {\langle {#1} \vert}

\begin{document}
\title{Near-linear Scaling in DMRG-based Tailored Coupled Clusters:\\
An Implementation of DLPNO-TCCSD and DLPNO-TCCSD(T)}

\author{Jakub Lang}
\email{jakub.lang@jh-inst.cas.cz}
\affiliation{J. Heyrovsk\'{y} Institute of Physical Chemistry, Academy of Sciences of the Czech \mbox{Republic, v.v.i.}, Dolej\v{s}kova 3, 18223 Prague 8, Czech Republic}
\affiliation{Faculty of Sciences, Charles University,  Albertov 6, 128 00 Praha 2, Czech Republic}

\author{Andrej Antal\'{i}k}
\email{andrej.antalik@jh-inst.cas.cz}
\affiliation{J. Heyrovsk\'{y} Institute of Physical Chemistry, Academy of Sciences of the Czech \mbox{Republic, v.v.i.}, Dolej\v{s}kova 3, 18223 Prague 8, Czech Republic}
\affiliation{Faculty of Mathematics and Physics, Charles University, Ke Karlovu 3, 12116, Prague 2, Czech Republic}

\author{Libor Veis}
\email{libor.veis@jh-inst.cas.cz}
\affiliation{J. Heyrovsk\'{y} Institute of Physical Chemistry, Academy of Sciences of the Czech \mbox{Republic, v.v.i.}, Dolej\v{s}kova 3, 18223 Prague 8, Czech Republic}

\author{Jan Brandejs}
\email{jan.brandejs@jh-inst.cas.cz}
\affiliation{J. Heyrovsk\'{y} Institute of Physical Chemistry, Academy of Sciences of the Czech \mbox{Republic, v.v.i.}, Dolej\v{s}kova 3, 18223 Prague 8, Czech Republic}
\affiliation{Faculty of Mathematics and Physics, Charles University, Ke Karlovu 3, 12116, Prague 2, Czech Republic}

\author{Ji\v{r}\'{i} Brabec}
\email{jiri.brabec@jh-inst.cas.cz}
\affiliation{J. Heyrovsk\'{y} Institute of Physical Chemistry, Academy of Sciences of the Czech \mbox{Republic, v.v.i.}, Dolej\v{s}kova 3, 18223 Prague 8, Czech Republic}

\author{\"Ors Legeza}
\email{legeza.ors@wigner.mta.hu}
\affiliation{Strongly Correlated Systems ``Lend\"{u}let'' Research group, Wigner Research Centre for Physics, H-1525, Budapest, Hungary}

\author{Ji\v{r}\'{i} Pittner}
\email{jiri.pittner@jh-inst.cas.cz}
\affiliation{J. Heyrovsk\'{y} Institute of Physical Chemistry, Academy of Sciences of the Czech \mbox{Republic, v.v.i.}, Dolej\v{s}kova 3, 18223 Prague 8, Czech Republic}

\date{\today}

\begin{abstract}
  We present a new implementation of density matrix renormalization group based tailored coupled clusters
  method (TCCSD), which employs the domain-based local pair natural orbital approach (DLPNO).
  Compared to the previous local pair natural orbital (LPNO) version of the method, the new implementation
  is more accurate, offers more favorable scaling and provides more consistent behavior across the variety
  of systems.
  On top of the singles and doubles, we include the perturbative triples correction (T), which is able to
  retrieve even more dynamic correlation.
  The methods were tested on three systems: tetramethyleneethane, oxo-Mn(Salen) and Iron(II)-porphyrin model.
  The first two were revisited to assess the performance with respect to LPNO-TCCSD.
  For oxo-Mn(Salen), we retrieved between 99.8--99.9\% of the total canonical correlation energy which is
  the improvement of 0.2\% over the LPNO version in less than 63\% of the total LPNO runtime.
  Similar results were obtained for Iron(II)-porphyrin.
  When the perturbative triples correction was employed, irrespective of the active space size or system,
  the obtained energy differences between two spin states were within the chemical accuracy of 1 kcal/mol
  using the default DLPNO settings.  
\end{abstract}


\maketitle

\section{Introduction}
\label{sec_intro}
Since its introduction to quantum chemistry \cite{Cizek1966}, the coupled cluster (CC) approach has become
one of the most widely used methods for the accurate calculations of dynamic correlation.
It offers numerous favorable properties, such as compact description of the wave function, size-extensivity,
invariance to rotations within occupied or virtual orbital subspaces and also a systematic hierarchy
of approximations converging towards the full configuration interaction (FCI) limit \cite{Gauss1998_encyclop}.
For instance, the CCSD(T) method \cite{Raghavachari1989}, which includes connected single-, double- and perturbative
triple excitations, is notorious referred to as the gold standard of quantum chemistry \cite{Gauss1998_encyclop}.

Although the CC method performs well for single reference molecules, it becomes fairly inaccurate or breaks down completely
for systems with strongly correlated electrons.
Such systems are multireference in nature since they include quasi-degenerate frontier orbitals.
This situation is common during dissociation processes, in diradicals, or compounds containing transition metals.
Over the years, numerous efforts to generalize the CC ansatz and thus overcome this drawback
gave rise to a broad family of multireference CC methods (MRCC) \cite{Bartlett2007,Tew2010,Lyakh2012}.

A possible approach for including static correlation in the CC scheme is to employ
a different method like complete active space self-consistent field (CASSCF) or multireference configuration
interaction (MRCI) in order to extract the information about the most important excitations
\cite{Li1997a,Li2001_theochem,Li2001_jcp_a,Li2001_jcp_b,Paldus1994,Piecuch1996,
Li1997b,Kinoshita2005,Lyakh2011,Melnichuk2012,Melnichuk2014,Piecuch1993,Piecuch1994,Adamowicz1998,Piecuch2010,
Piecuch2002_1,Piecuch2002_2,Kowalski2002,Wloch2006,Lodriguito2006,Piecuch2004,Kowalski2001,Kowalski2000}.
The retrieved information can then be introduced to a CC calculation as an external correction.
One of such methods is tailored CC with single and double excitations (TCCSD) proposed by
Kinoshita et al. \cite{Kinoshita2005},
which draws on the split-amplitude ansatz by Piecuch et al.\cite{Piecuch1993},
in which the cluster operator corresponding to single and double
excitations is split into two parts.
The active part is imported from a complete active space configuration interaction (CAS-CI) and kept fixed,
while the external amplitudes are iterated using the standard CCSD framework.
We recently extended this approach by using the density matrix renormalization group (DMRG)
method to obtain the active space amplitudes \cite{Veis2016,Veis2016_corr}.
Related externally corrected CC approaches employ fixed $T_3$ and $T_4$ amplitudes obtained from $MRCI$ \cite{paldus-externalcorr} or from stochastic CI \cite{piecuch-ccmc}
and iterate all singles and doubles in their presence. Compared to TCC, this has the advantage that the active space  $T_1$ and $T_2$ can reflect the dynamic correlation
outside of the active space, but one pays the price of much larger number of the  $T_3$ and $T_4$ amplitudes involved.

The DMRG method, which originated in solid-state physics \cite{White1992a,White1992b,White1993}, is nowadays well established
in quantum chemistry for the treatment of strongly correlated systems
\cite{White1999,Chan2002a,Legeza2003a,Legeza2008,Marti2010c,Chan2011,Wouters2014_rev,Szalay2015,Yanai2014}.
As a numerical approximation to FCI, it can handle significantly
larger active spaces compared to the conventional methods.
However, even then the prohibitive scaling does not allow to include dynamic correlation
and it is therefore necessary to employ some "post-DMRG" procedure.
Many different attempts has been made to tackle this limitation 
for example with
the complete active space perturbation theory (CASPT2)\cite{Kurashige2011},
Cholesky decomposition n-electron valence state perturbation theory (NEVPT2)\cite{Freitag2017},
MRCI using cumulant reconstruction with internal contraction of DMRG wave function\cite{Saitow2013},
canonical transformation\cite{Neuscamman2010},
matrix product state (MPS) based formulation of multireference perturbation theory\cite{Sharma2014},
DMRG pair-density functional theory\cite{Sharma2019},
and also our aforementioned CC tailored by MPS wave functions (DMRG-TCCSD)\cite{Veis2016,Faulstich2019,Faulstich2019_n2}.

Even though the DMRG-TCCSD method offers a reasonably efficient treatment of both static and dynamic
correlation\cite{Veis2018}, its application to larger systems is hampered by the unfavorable scaling
of the CCSD part of the calculation.
With such a steep scaling, even massive parallelization is not sufficient to make the method
applicable to molecules with hundreds of atoms.
To overcome this issue, Pulay proposed to exploit the locality of the electron
correlation\cite{Pulay1983,Saebo1985}.
Due to its short range character for non-metallic systems, it is possible to take advantage of
sparsity of the Hamiltonian matrix by employing the basis of localized orbitals.

Although the occupied orbital space can be easily localized using a variety of appropriate methods
\cite{Foster1960,Pipek1989,Knizia2013}, for the virtual space things get slightly
more complicated.
In their first works on locality, Pulay and S{\ae}b{\o} used projected atomic orbitals (PAOs)
\cite{Saebo1987,Saebo1988}, which were also used by Werner and Sch\"{u}tz in the local CC method
\cite{Hampel1996,Schutz2001,Schutz2002b,Werner2011,Schutz2002a}.
In this approach, each localized occupied orbital is assigned a domain of PAOs obtained
by projecting out the occupied orbital components from atomic orbitals.
The pairs of occupied orbitals are subsequently classified according to their real space distance
and treated at either coupled cluster level (strong pairs), perturbative level (weak and distant pairs),
or neglected altogether (very distant pairs).

Another idea how to make use of locality is based on the concept of dividing a large system into
smaller segments and performing the calculations on each of these subsystems separately.
Among such approaches belong the divide-expand-consolidate method \cite{Kristensen2011,Hoyvik2012},
the divide-and-conquer method \cite{Kobayashi2008}, the incremental method \cite{Stoll1992},
the local natural orbital method \cite{Rolik2011}, and the fragment molecular orbital method \cite{Fedorov2005}.
The closely related cluster-in-molecule method \cite{Li2001} is based on energy decomposition
into contributions corresponding to individual occupied orbitals.

Possibly the most effective way of virtual space truncation is, however, the use of pair natural orbitals (PNOs),
which are known to provide compact parametrization of the virtual space.
They were first used in the 1960s by Edmiston and Kraus \cite{Edmiston1965} and later
by Meyer \cite{Meyer1971,Meyer1974,Werner1976,Botschwina1977,Rosmus1978}, Ahlrichs and Kutzelnigg \cite{Ahlrichs1968,Ahlrichs1975}.
For many years, the progress in the area of PNO-based methods stalled, until their revival in 2009, when the new
local pair natural orbital (LPNO) variants of CEPA and CCSD methods were
introduced by Neese et al\cite{Neese2009_cepa,Neese2009_ccsd,Hansen2011,Huntington2012}.
The cornerstone of this approach, the idea to use PNOs in combination with localized occupied orbitals, was
later developed into the more advanced domain based local pair natural orbital (DLPNO)
methods\cite{Riplinger2013_dlpno,Pinski2015,Riplinger2016,Saitow2017}.

In these, PNOs were expressed as a linear combination of PAOs in a pair-domain, which ultimately
removed the bottleneck of previous PNO methods and achieved genuine linear scaling.
For instance, the resulting DLPNO-CCSD method is applicable to systems with hundreds of atoms and
thousands of basis functions, which renders the prior SCF calculation possibly computationally more
demanding than the actual correlation treatment.

Moreover, the PNO-based approaches possess many desirable properties, which allow them to be used in
black box fashion.
They provide a very compact description of the virtual space, which makes it computationally feasible
to use sufficiently large domains of PAOs; something that would be too costly for a purely PAO-based approach.
They use a limited number of cut-off parameters, which do not involve any real space distance and
the calculated correlation energy smoothly depends on the values of these parameters.

Currently, PNO-based methods are developed in a number of groups including
Werner\cite{Werner2015,Menezes2016,Schwilk2017,Tew2011}
and H\"{a}ttig \cite{Helmich2013,Schmitz2013,Guo2016}
and are widely employed in the context of various systems of chemical
interest\cite{Antony2011,Anoop2010,Liakos2011,Zade2011,Kubas2012,Ashtari2011,Zhang2014,Minenkov2015,Sparta2014,Liakos2015}.
Apart from single-reference methods, the methodology was also successfully applied
to multireference CC techniques \cite{Demel2015,Lang2017,Brabec2018,Lang2019}.

In this article, we build upon the previous investigation of our LPNO-TCCSD method\cite{Antalik2019}
and introduce the newly implemented DLPNO-TCCSD and DLPNO-TCCSD(T) methods.
Based on our experience, we revisit the molecule of tetramethyleneethane (TME) and address the
drawbacks of the former method.
The performance of the new approximation is then assessed using two benchmark systems, namely,
oxo-Mn(Salen) and a model of Fe(II)-porphyrins (FeP).

\section{Theory and Implementation}
\label{sec_theory}

\subsection{DMRG-based Tailored Coupled Clusters}

The tailored coupled cluster method\cite{Kinoshita2005}, which belongs to the class of externally corrected methods, employs
the split-amplitude wave function ansatz proposed by Piecuch et al. \cite{Piecuch1993}
\begin{equation}
  | \Psi_\text{TCC} \rangle = e^T | \Phi_0 \rangle
                            = e^{T_\text{ext}+T_\text{CAS}} | \Phi_0 \rangle
                            = e^{T_\text{ext}}e^{T_\text{CAS}} | \Phi_0 \rangle,
\end{equation}
where $| \Phi_0 \rangle$ is the reference wave function and the cluster operator $T$ is split into two parts:
$T_\text{CAS}$ which contains the active amplitudes obtained from an external calculation and
$T_\text{ext}$ which contains the external amplitudes i.e. the amplitudes with at least one index
outside the CAS space.
Another way to justify this ansatz is the formulation of CC equations based on excitation subalgebras
recently introduced by Kowalski \cite{Kowalski2018,Bauman2019}.

In our implementation, we employed the DMRG method to obtain the active amplitudes.
Using the DMRG algorithm we first optimize the wave function, which is provided in
the MPS form
\begin{equation}
  | \Psi_\text{MPS} \rangle = \sum_{\{\alpha\}}
  \mathbf{A}^{\alpha_1} \mathbf{A}^{\alpha_2} \cdots \mathbf{A}^{\alpha_k}
  | \alpha_1 \alpha_2 \cdots \alpha_k \rangle,
\end{equation}
where $\alpha \in \{ |-\rangle, |\downarrow\rangle, |\uparrow\rangle, |\downarrow\uparrow\rangle \}$
and $\mathbf{A}^{\alpha_i}$ are MPS matrices.
These are then contracted to obtain CI coefficients for single and double excitations $C$
\cite{Moritz2007,Boguslawski2011}.
Using the intermediate normalization and the relations between CI and CC coefficients
\begin{align}
  T^{(1)}_\text{CAS} &= C^{(1)}, \\
  T^{(2)}_\text{CAS} &= C^{(2)} - \frac{1}{2}[C^{(1)}]^2,
\end{align}
we are able to acquire the respective CC amplitudes, which are subsequently introduced into
the CC calculation.
At this point, these active amplitudes are kept frozen, while the remaining amplitudes in $T_\text{ext}$
are optimized by solving the equations
\begin{align}
  \langle \Phi_i^a | He^{T_\text{ext}}e^{T_\text{CAS}} | \Phi_0 \rangle_c &= 0 \quad
  \{ i,a \} \not\subset \text{CAS} \label{eq_TCC_singles} \\
  \langle \Phi_{ij}^{ab} | He^{T_\text{ext}}e^{T_\text{CAS}} | \Phi_0 \rangle_c &= 0 \quad
  \{ i,j,a,b \} \not\subset \text{CAS} \label{eq_TCC_doubles}
\end{align}
analogously to the standard CCSD equations.
This way, the active amplitudes account for static correlation and by optimizing the external amplitudes, we are able
to recover the remaining dynamic correlation.
The effect of dynamic correlation on $T_\text{CAS}$ is neglected, which is an approximation inherent in
the TCC method.

Due to the two-body Hamiltonian, TCC recovers the DMRG energy for $T_\text{ext}=0$.
In the limit of CAS including all orbitals, FCI energy is recovered, although the TCC energy
does not behave monotonously when extending the CAS space\cite{Faulstich2019_n2}.
Nevertheless, in practice the optimal CAS size related to the energy minimum can be determined with
low cost DMRG calculations\cite{Faulstich2019_n2}.
In addition, a quadratic error bound valid for DMRG-TCC methods is also derived\cite{Faulstich2019_n2}.

On top of the TCCSD routine, the perturbative triples correction can be applied\cite{Lyakh2011}.
However, in order to prevent the double counting of static correlation, it is necessary
to omit the terms that include the active amplitudes.
This can be straightforwardly achieved by setting all active single and double amplitudes to zero during
the calculation of the (T) correction.

\subsection{Domain Local Pair Natural Orbital Approximation for TCCSD}
The presented method is based on the open-shell DLPNO-CCSD code as implemented in ORCA\cite{Neese2011}.
In this section, we briefly outline DLPNO-CCSD and describe the modifications that
were made to accommodate the tailored version of the method.

As in all PNO-based local methods, the whole process starts with the localization of the occupied
orbitals.
Based on our previous experience\cite{Antalik2019}, we opt for the split-localization scheme, where the
orbitals are separately localized within four distinct orbital subspaces: doubly occupied external,
doubly occupied active, singly occupied active and active virtual.
Such choice has been shown to provide a set of orbitals which yields a reasonable convergence behavior
of the DMRG procedure\cite{OlivaresAmaya2015}, without influencing the DMRG energy.

The next step is the construction of orbital domains.
Using the idea of Werner et al.\cite{Adler2011,Adler2009}, we first construct PAOs
\begin{equation}\label{PAO}
  \ket{\tilde\mu} = \Big( 1 - \sum_{i} \ket{i} \bra{i} \Big) \ket{\mu} 
\end{equation}
by projecting out the localized occupied and active orbitals $|i\rangle$ from the original set of
atomic orbitals $|\mu\rangle$.
The acquired orbitals are subsequently normalized and the orbital domains are constructed
based on the differential overlap integrals
\begin{equation}\label{DOI}
  \text{(DOI)}_{i\mu} = \sqrt{\int_{\mathbb{R}^3} |\phi_i|^2 |\phi_{\tilde\mu}|^2 \diff^3r}
\end{equation}
between PAOs and the set of occupied and active orbitals.
Here, the first prescreening parameter comes into play -- for a given occupied orbital $i$,
only PAOs for which $\text{(DOI)}_{i\mu} > T_\text{CutDO}$ are included within its domain.
Also, if an atom contains at least one PAO, all PAOs belonging to this atom are considered for
further domain construction.

At this point, the main difference of the TCC approach compared to the conventional DLPNO-CCSD is
that all active indices including the active virtuals are formally treated as singly occupied.
This means that during the creation of domains they share the same domain.
Moreover, in the following dipole prescreening, it is ensured that every active occupied
pair (i.e. a pair with both active indices) automatically survives the dipole prescreening.

After the prescreenings, unrestricted MP2 pair energies $\varepsilon_{ij}^\text{PAO}$ are calculated
and used to categorize occupied orbital pairs $ij$ into three classes, based on the preset cut-off parameters.
Specifically, the active pairs and the pairs with energies larger than $T_\text{CutPairs}$ are
classified as strong, while the rest is either weak or neglected according to a related parameter.
The whole process is executed in two consecutive steps.
First, a crude screening is performed in smaller domains using loose thresholds, followed by the
second screening in which the strong and weak pairs are again distributed between the three
categories, but this time with finer thresholds.
The strong pairs are then passed to the next stage, while the remaining pair energies are stored as
a correction to the final energy
\begin{equation}
  \Delta E_\text{CutPairs} = \sum_{ij}^\text{weak}\varepsilon_{ij}^\text{PAO} \\
    + \sum_{ip}^\text{weak}\varepsilon_{ip}^\text{PAO} + \sum_{pq}^\text{weak}\varepsilon_{pq}^\text{PAO},
  \label{dE_Pairs}
\end{equation}
where $i$, $j$ are indices of doubly and $p$, $q$ of singly occupied orbitals.
If the perturbative triples correction is invoked, the final weak pairs with energies higher than
$0.01 \cdot T_\text{CutPairs}$ are also saved for later use.

\begin{figure}
  \centering
  \includegraphics[width=0.35\textwidth]{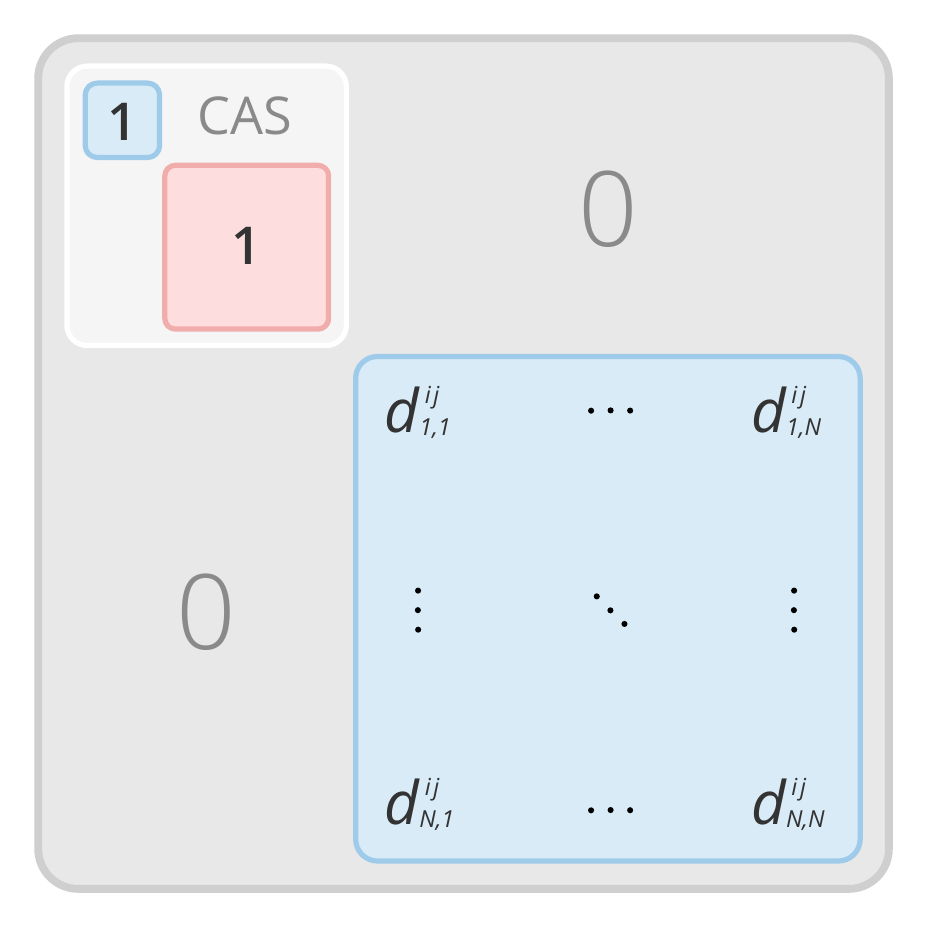}
  \caption{An illustration of the PAO/PNO transformation matrix for an active pair $ij$ in DLPNO-TCC.
  The original transformation matrix $\mathbf{d}^{ij}$, with $N$ being the number of PNOs, is enlarged
  by an identity matrix of size $N_\text{CAS}$, which is formally composed of two blocks corresponding
  to singly occupied (blue) and virtual orbitals (red) included in the active space.}
  \label{fig:fig_pdm}
\end{figure}

Afterwards, using the non-redundant PAOs, the NEVPT2 pair densities $\mathbf{D}^{ij}$ are constructed for the
surviving pairs which do not contain any explicit information about the tailored CAS space.
These are diagonalized to obtain PNO expansions $\mathbf{d}^{ij}$, which are then truncated
based on the cut-off parameter $T_\text{CutPNO}$.
Only PNOs with occupation numbers larger than its value are kept and the remaining orbitals are discarded.
The final PAO/PNO transformation matrix for a given pair is then obtained by enlarging the former
transformation matrix by a unit matrix
\begin{equation}
  \mathbf{S}^{ij} = \mathbf{I}_{N_\text{CAS}} \oplus \mathbf{d}^{ij},
\end{equation}
where $N_\text{CAS}$ is the number of active SOMOs and virtual orbitals, see Figure \ref{fig:fig_pdm}.
While this artificially increases the final number of PNOs, it barely increases the computational time.
Singles PNOs are obtained in the same manner from the densities which are computed as
\begin{equation}\label{singlesPNO}
  \mathbf{D}^{pp}=\mathbf{t}^{(1)\dagger}_{pp} \mathbf{t}^{(1)}_{pp},
\end{equation}
where the amplitudes are
\begin{equation}
  \mathbf{t}^{(1)}_{pp} = \sum_{\tilde\mu_{pp}\tilde\nu_{pp}}
  \frac{ \bra{p\tilde\mu_{pp}} \ket{p\tilde\nu_{pp}} }
  {2f_{pp}-\varepsilon_{\tilde\mu_{pp}}-\varepsilon_{\tilde\nu_{pp}}},
\end{equation}
where $f_{pp}$ are diagonal Fock matrix elements.
Furthermore, the pair energies are recalculated in the new truncated PNO basis and the differences between them
and the original MP2 estimates in the PAO basis
\begin{equation}
  \begin{split}
    \Delta E_\text{CutPNO} = &\sum_{ij}^\text{strong}(\varepsilon_{ij}^\text{PAO}-\varepsilon_{ij}^\text{PNO}) \\
    + &\sum_{ip}^\text{strong}(\varepsilon_{ip}^\text{PAO}-\varepsilon_{ip}^\text{PNO})\\
    + &\sum_{pq}^\text{strong}(\varepsilon_{pq}^\text{PAO}-\varepsilon_{pq}^\text{PNO}),
  \end{split}
  \label{dE_PNO}
\end{equation}
are stored as a correction to the final energy.

The resulting equations for singly excited amplitudes (\ref{eq_TCC_singles}) now become
\begin{equation}
  \langle \Phi_{i}^{\bar a} | He^{\bar T^{(1)}_\text{ext} + \bar T^{(2)}_\text{ext}}e^{T_\text{CAS}}
  | \Phi_0 \rangle_c = 0 \quad \{ i,a \} \not\subset \text{CAS},
\end{equation}
where the barred index $\bar a$ indicates PNO basis.
Similarly, the equations for doubly excited amplitudes (\ref{eq_TCC_doubles}) become
\begin{equation}
  \langle \Phi_{ij}^{\bar a \bar b} | He^{\bar T^{(1)}_\text{ext} + \bar T^{(2)}_\text{ext}}e^{T_\text{CAS}}
  | \Phi_0 \rangle_c = 0 \quad \{ i,j,a,b \} \not\subset \text{CAS}.
\end{equation}
Once these equations are solved, the stored corrections (\ref{dE_Pairs}) and (\ref{dE_PNO}) are added
to the acquired energy.
Except the fact that the active amplitudes are `frozen' during the CCSD iterations, these equations
are identical to single-reference DLPNO-CCSD as implemented in ORCA\cite{Saitow2017}.

\subsection{Perturbative triples correction to DLPNO-TCCSD}

To calculate the triples correction using the DLPNO approach\cite{Riplinger2013_triples,Guo2018}, it is first necessary
to identify relevant compact parameterization of virtual space for every triple $ijk$.
Domains for these triples are created as a union of $i$, $j$ and $k$ domains and the 
$ij$, $jk$ and $ik$ pair densities are created within these triples domains.

Next, the triples densities are constructed by averaging over the respective pair densities
\begin{equation}\label{key}
  \mathbf{D}^{ijk}=\frac{1}{3}\Big(\mathbf{D}^{ij} + \mathbf{D}^{jk} + \mathbf{D}^{ik}\Big),
\end{equation}
where the triple $ijk$ is composed either of three strong pairs or of two strong and a weak pair, as it has
been previously shown that using only the strong pairs is insufficient\cite{Riplinger2013_triples}.
Since the amplitudes for weak pairs are not known prior to the triples calculation, the approximate
MP2 amplitudes are used.
The process then continues analogously to the construction of PNOs in CCSD.

Once the densities are constructed, they are diagonalized in order to obtain the triple natural orbitals (TNO)
and their corresponding natural occupation numbers.
The TNO expansion is then truncated based on the occupation numbers and the cut-off parameter
$T_\text{CutTNO}$ and from the orbitals that passed the screening a transformation matrix is formed.
This matrix is enlarged by a unit matrix of dimension $N_\text{CAS}$ and its final form is used to transform
the integrals, amplitudes and subsequently to calculate the energy correction.
As in the canonical version of the method, all active single and double amplitudes are set to zero during
the calculation of the correction to prevent double-counting.


%

\begin{figure*}
  \centering
  \includegraphics[width=0.9\textwidth]{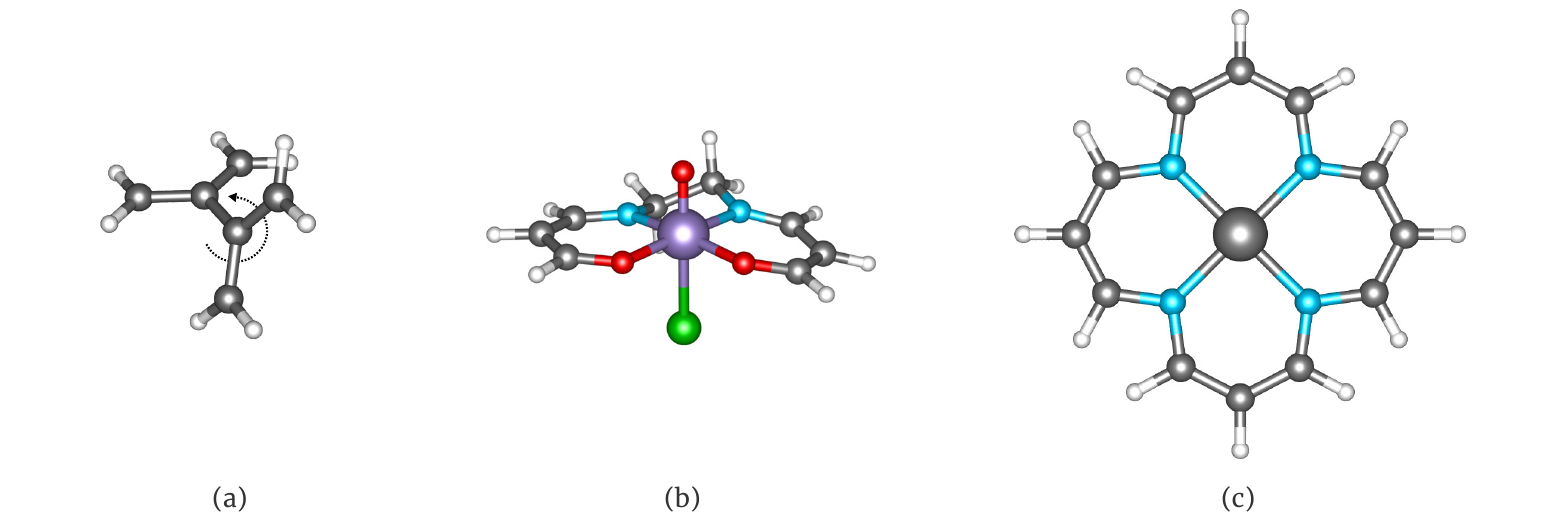} 
  \caption{Dihedral rotation of tetramethyleneethane (a), a molecule of oxo-Mn(Salen) (b)
  and the model of Fe(II)-porphyrin (c).
  Color key: iron (gray, large), manganese (violet), chlorine (green), oxygen (red), nitrogen (blue),
  carbon (gray, small), hydrogen (white). \label{fig:systems}}
\end{figure*}

\section{Computational Details}
\label{sec_compdet}
The DMRG calculations were performed by the Budapest QC-DMRG code \cite{budapest_qcdmrg}.
The DLPNO-TCCSD and DLPNO-TCCSD(T) methods were implemented in the ORCA program package \cite{Neese2011},
which was also used to prepare the orbitals.

Prior to DMRG calculations, we split-localized the orbitals by the Pipek-Mezey method\cite{Pipek1989}
in the following orbital subspaces: internal, active doubly occupied, active singly occupied and active virtual.
Orbital ordering was subsequently optimized using the Fiedler method \cite{Barcza2011, Fertitta2014}
combined with minor manual adjustments.
All DMRG runs were initialized by the CI-DEAS procedure \cite{Legeza2003,Szalay2015} and employed
the dynamical block state selection (DBSS) procedure \cite{Legeza2003a, Legeza2004} to control
the accuracy of the calculation, with the truncation error criterion set to $10^{-6}$.
The convergence threshold was set to energy difference between two subsequent sweeps smaller than $10^{-6}$ a.u.

The core electrons were kept frozen throughout all coupled cluster calculations.
The default set of DLPNO cut-off parameters
$T_\text{CutPNO}=3.33\cdot10^{-7}$, $T_\text{CutPairs}=10^{-4}$, $T_\text{CutMKN}=10^{-3}$ and
$T_\text{CutDO}=10^{-2}$ was employed unless otherwise stated.
The other settings, referred to as TightPNO, were used for production runs:
$T_\text{CutPNO}=10^{-7}$, $T_\text{CutPairs}=10^{-5}$, $T_\text{CutMKN}=10^{-3}$ and
$T_\text{CutDO}=5.0\cdot10^{-3}$.
For perturbative triples the default value of relevant cut-off parameter is $T_\text{CutTNO}=10^{-9}$
for both sets of settings.
To estimate the dependence of DLPNO-TCCSD energies on these parameters, one parameter was varied with
remaining parameters fixed to the default value.
We assess the amount of retrieved correlation energy by DLPNO approach with reference to the DMRG-TCCSD energy
calculated with the canonical TCCSD implementation.

For TME, we used CASPT2(6,6)/cc-pVTZ geometries for seven values of the dihedral angle from our
previous work \cite{Veis2018}.
The orbitals were prepared by CASSCF(6,6) calculation with the active space containing
six 2p$_z$ orbitals on carbon atoms.
The cc-pV6Z/C auxiliary basis set was used for the resolution of the identity (RI) approximation \cite{Bross2013}.

For oxo-Mn(Salen), we used the singlet CASSCF(10,10)/6-31G* optimized geometry by Ivanic et al. \cite{Ivanic2004}.
The orbitals were optimized using the DMRG-CASSCF method \cite{Ghosh2008,Zgid2008,Yanai2009} in Dunning's cc-pVXZ
$\text{X}\in$\{D,T,Q\} basis sets \cite{Dunning1989,Woon1993,Balabanov2005}.
The optimization was carried out with fixed bond dimension $M=1024$ for the smaller CAS(28,22)
and $M=2048$ for CAS(28,27).
The composition of these active spaces is discussed further in detail in the paper\cite{Antalik2019}.
The cc-pVQZ/C auxiliary basis set was used for RI \cite{Weigend2002}.

Finally, the calculations on FeP were performed at the triplet geometry from the previous study 
of Li Manni and Alavi \cite{LiManni2018},
with CASSCF/def2-SVP orbitals, for three distinct active spaces, which were selected
based on the entanglement analysis \cite{}.
For the largest CAS the DMRG-CASSCF was applied, with the fixed bond dimension $M=1024$.
The def2-SVP and def2-TZVP basis sets\cite{Weigend2005} were used with the def2-TZVPP/C auxiliary
basis set\cite{Hellweg2007} for RI.

\section{Results and Discussion}
\label{sec_results}

\subsection{Tetramethyleneethane}
Although small, the tetramethyleneethane molecule is a challenging system due to its complex electronic structure.
To correctly describe the character of its singlet state, one needs to employ a theory with a balanced description of
both static and dynamic correlation combined with a reasonably large basis set.
This is the reason, why it often serves as a benchmark system for multireference
methods \cite{Pittner2001, BhaskaranNair2011, Chattopadhyay2011, Pozun2013, Demel2015, Veis2018}.
We previously studied the system with the canonical DMRG-TCCSD method \cite{Veis2018}, as well as 
with its LPNO version\cite{Antalik2019}.
In the latter, we encountered an issue with different accuracy for singlet and triplet states, which
we attributed to the neglect of some terms in the LPNO-CCSD implementation in ORCA.
For this reason, we revisit the system to investigate the improvement by the new DLPNO method,
which should offer more robust approximation.

We performed calculations in seven different geometries corresponding to the rotation about the central
C--C bond of the molecule (see Figure \ref{fig:systems}a) and different values of the cut-off parameters.
These benchmarks were performed only for $T_\text{CutPairs}$ and $T_\text{CutPNO}$ parameters,
since $T_\text{CutDO}$ does not affect the results, unless an extremely small value is chosen.
This behavior corresponds with the observations from previous studies\cite{Lang2019,Brabec2018,Saitow2017}.

\begin{figure*}
  \centering
  \includegraphics[width=0.48\textwidth]{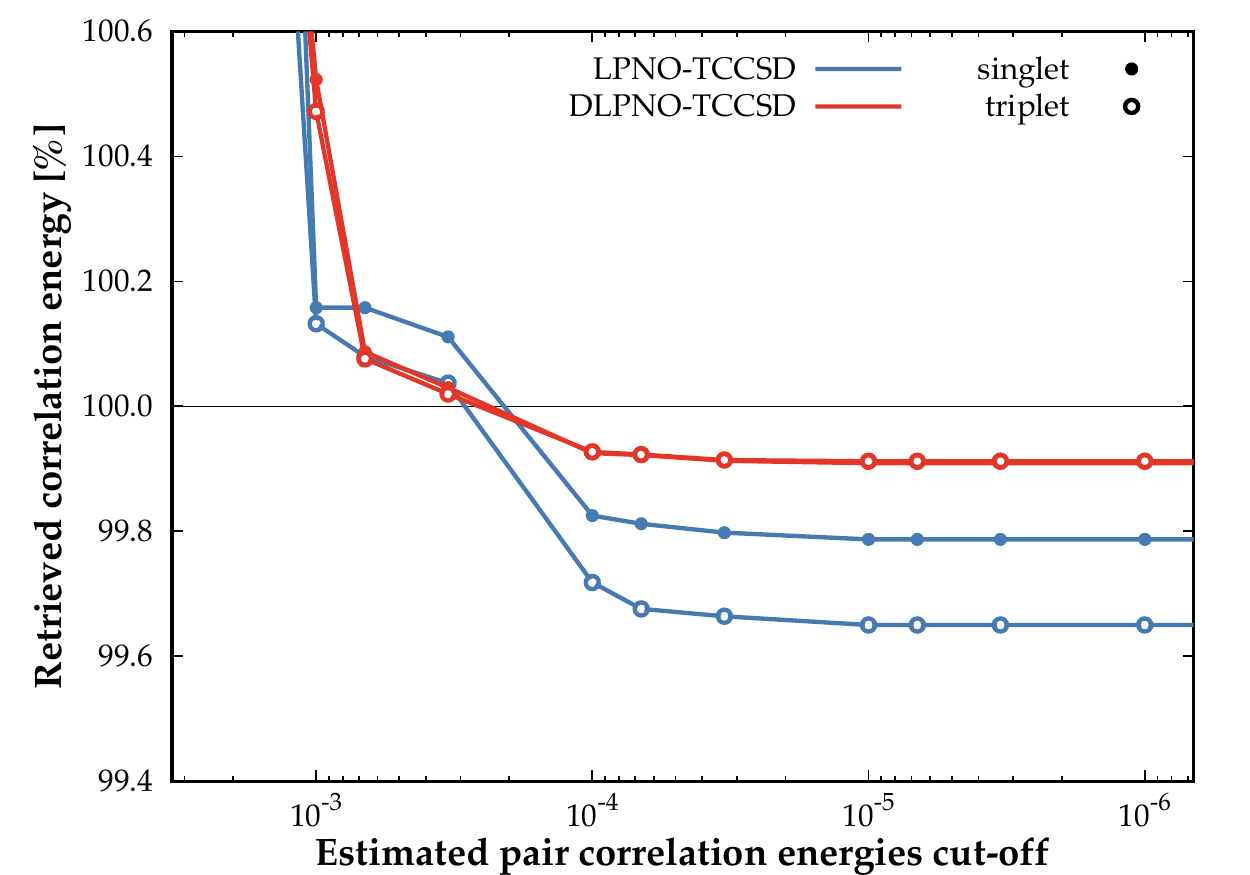}
  \hfill
	\includegraphics[width=0.48\textwidth]{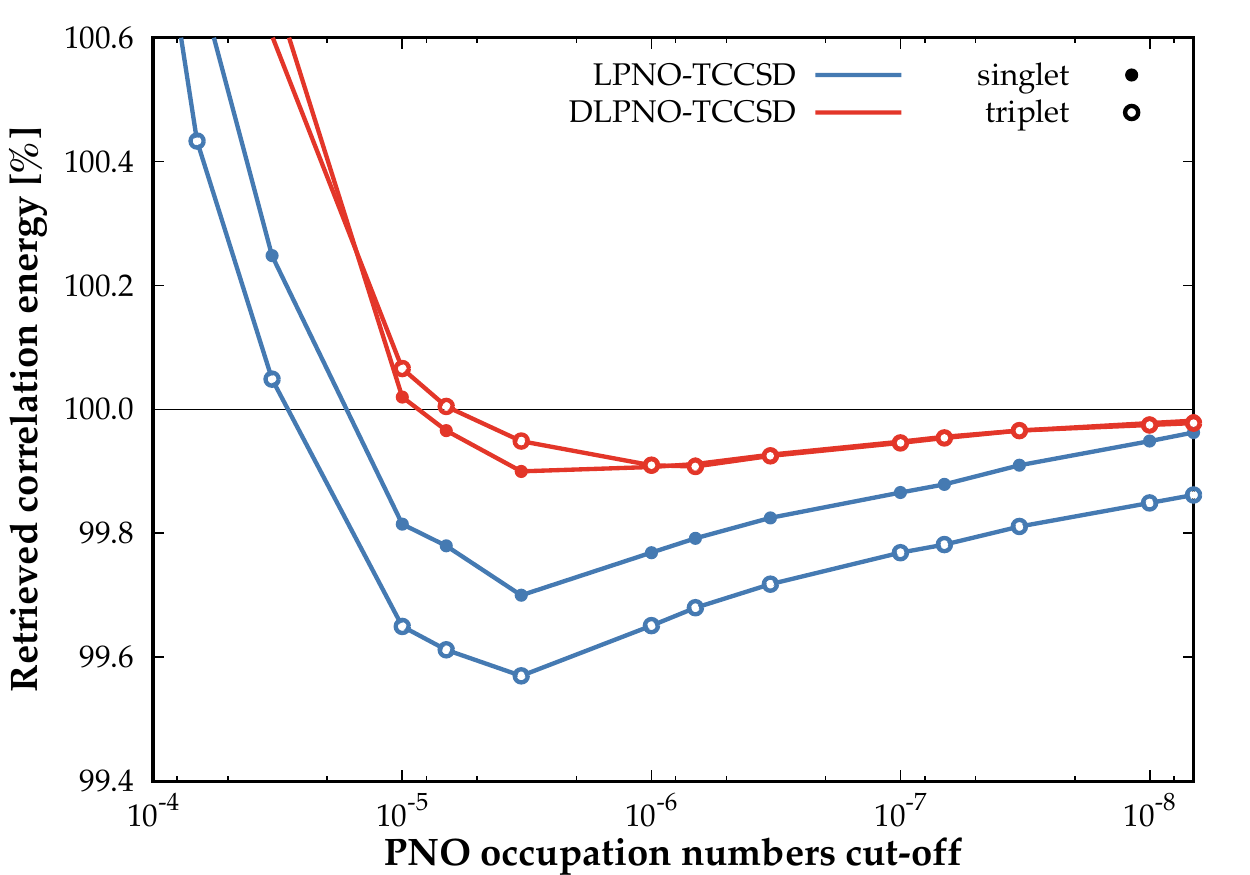}
  \caption{The percentage of correlation energy of TME in cc-pVTZ basis retrieved by LPNO-TCCSD and DLPNO-TCCSD
  with respect to canonical TCCSD calculations as a function of cut-off for estimated pair correlation
  energies $T_\text{CutPairs}$ (left) and PNO occupation numbers $T_\text{CutPNO}$ (right).
  The plotted values are averages over the set of all geometries.}
  \label{fig:TME_cutoffs}
\end{figure*}

The left plot in Figure \ref{fig:TME_cutoffs} shows the dependence of retrieved canonical correlation energy
with respect to estimated pair correlation energies cut-off $T_\text{CutPairs}$ averaged over the geometries.
Both methods converge in a similar fashion, but DLPNO-TCCSD recovers 0.1--0.3\%  more
correlation energy than LPNO-TCCSD.
The right plot in the same figure shows the dependence on the second cut-off parameter, that is PNO occupation
number $T_\text{CutPNO}$, which is again averaged over the geometries.
Here, the DLPNO-TCCSD shows noticeably faster convergence to the canonical value compared to LPNO-TCCSD,
with more accurate correlation energies even for higher values of the parameter.
For instance, when the default values are used, DLPNO-TCCSD extracts over 99.91\% of the canonical
correlation energy.

One can notice the aforementioned discrepancy in accuracy for LPNO-TCCSD method, which can be explained by
the neglected terms in the UHF-LPNO formalism.
DLPNO-TCCSD does not exhibit this behavior and describes both spin states equally well.
This means that the difference of the recovered correlation energy is less than 0.01\% compared
to LPNO-TCCSD, for which this percentage is an order of magnitude worse.

\begin{figure}
	\centering
	\includegraphics[width=0.48\textwidth]{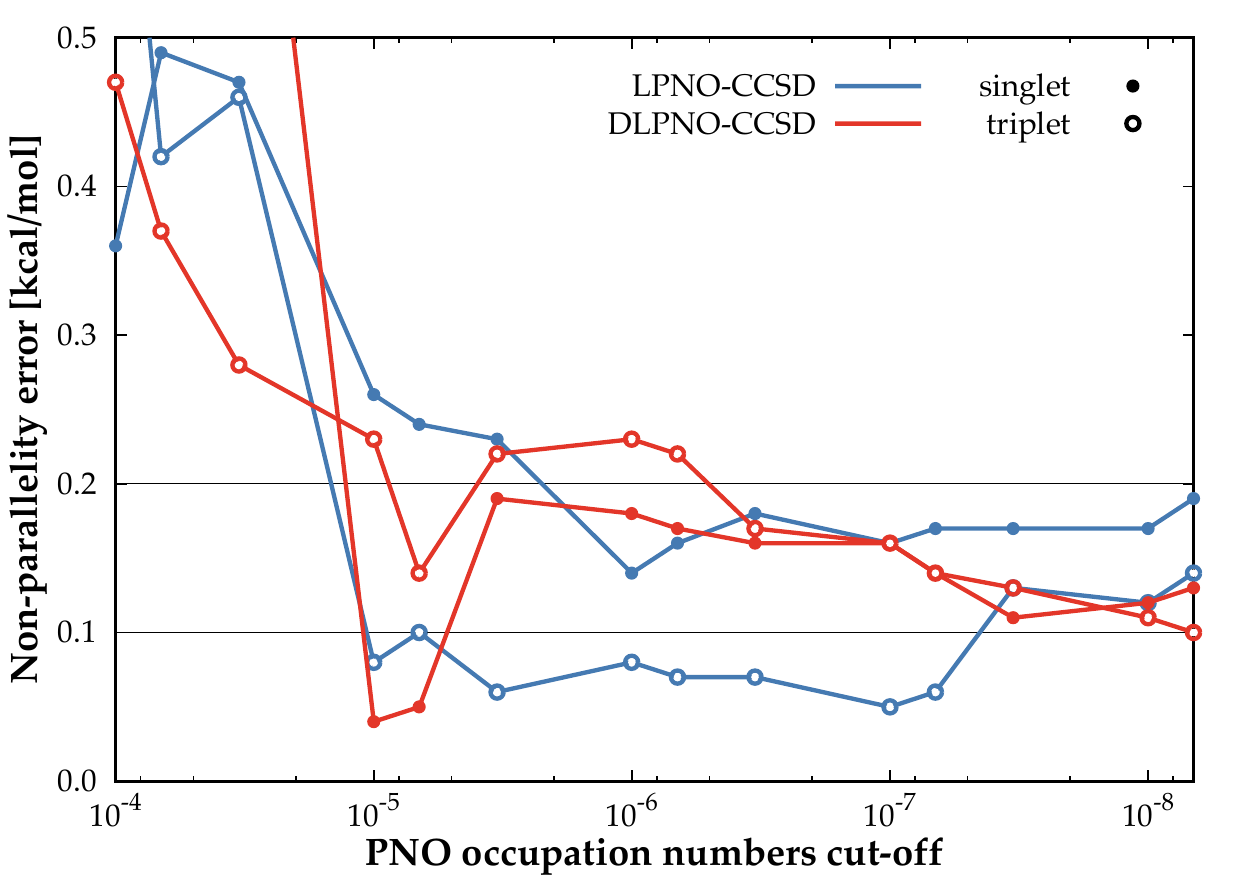}
	\caption{Non-parallelity error for TME in cc-pVTZ as a function of cut-off for PNO occupation numbers
  $T_\text{CutPNO}$.}
  \label{fig:TME_npe}	
\end{figure}

Figure \ref{fig:TME_npe} shows the dependence of non-parallelity error (NPE) on $T_\text{CutPNO}$ for both
LPNO and DLPNO version of TCCSD.
It can be seen that the NPE is no larger than 0.17 kcal/mol for either spin state, with very similar
value for TightPNO settings, as well as for the LPNO version.

From the chemical point of view, the most interesting property is the behavior of the singlet-triplet gap
with respect to the cut-off parameters.
It can be observed from the presented data that this property is well described at the default and even
looser settings, which means that it can be calculated by DLPNO-TCCSD with virtually no loss in accuracy.

\subsection{oxo-Mn(Salen)}
The oxo-Mn(Salen) molecule (Figure \ref{fig:systems}b) has been a subject of numerous computational studies
motivated mainly by its role in catalysis of the enantioselective epoxidation of unfunctional olefines
\cite{Zhang1990,Irie1990}.
Due to its closely lying singlet and triplet states it is also considered to be a challenging system
even for multireference methods.
Over the years, many multireference studies have been published \cite{Ivanic2004,Sears2006,Ma2011}, some
of which employed the DMRG method \cite{Wouters2014_oxo,OlivaresAmaya2015,Stein2016} and DMRG results
with dynamic correlation treatment were presented as well\cite{Veis2016,Sharma2017,Antalik2019}.
This includes our previous studies, in which we used the system as a benchmark for the predecessors
of the current method, namely DMRG-TCCSD and its LPNO version.

\begin{figure*}
  \centering
  \includegraphics[width=0.48\textwidth]{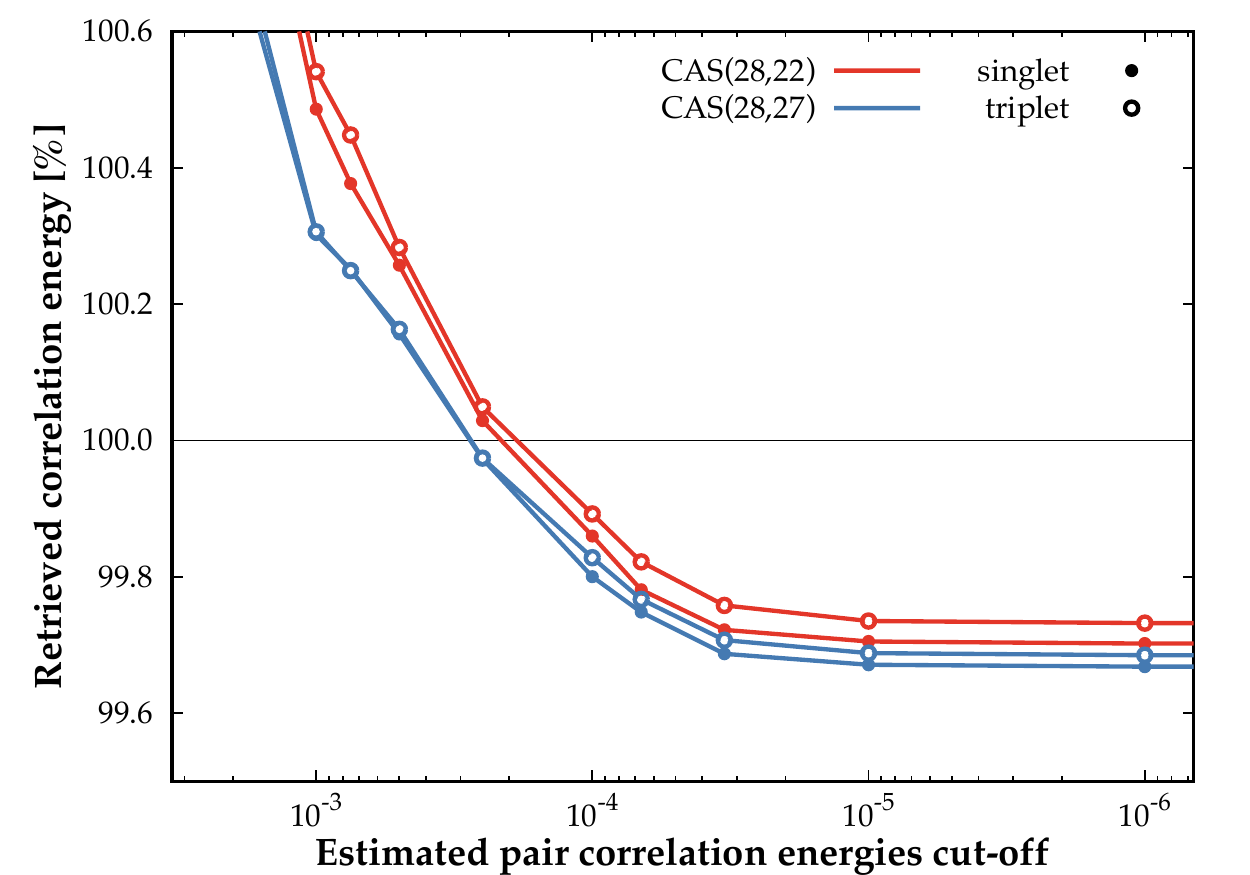}
  \hfill
  \includegraphics[width=0.48\textwidth]{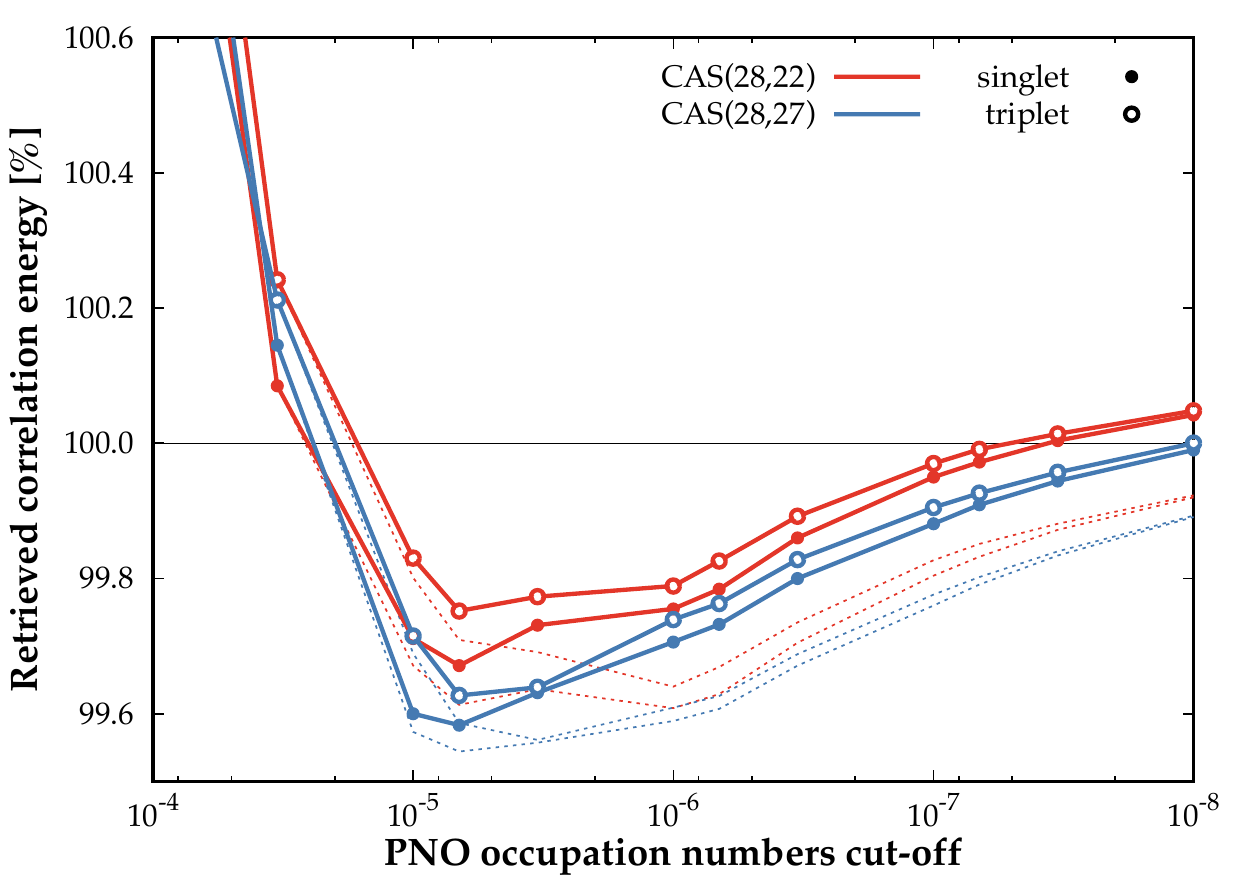}
  \caption{The percentage of correlation energy of oxo-Mn(Salen) in cc-pVDZ basis retrieved by DLPNO-TCCSD
  with respect to canonical TCCSD calculations as a function of cut-off for estimated pair correlation
  energies $T_\text{CutPairs}$ (left) and cut-off for PNO occupation numbers $T_\text{CutPNO}$ (right).
  The thin dashed lines in the right plot represent the values for $T_\text{CutPairs} = 10^{-5}$.}
  \label{fig:oxo_SD}
\end{figure*}

\begin{figure*}
  \centering
  \includegraphics[width=0.48\textwidth]{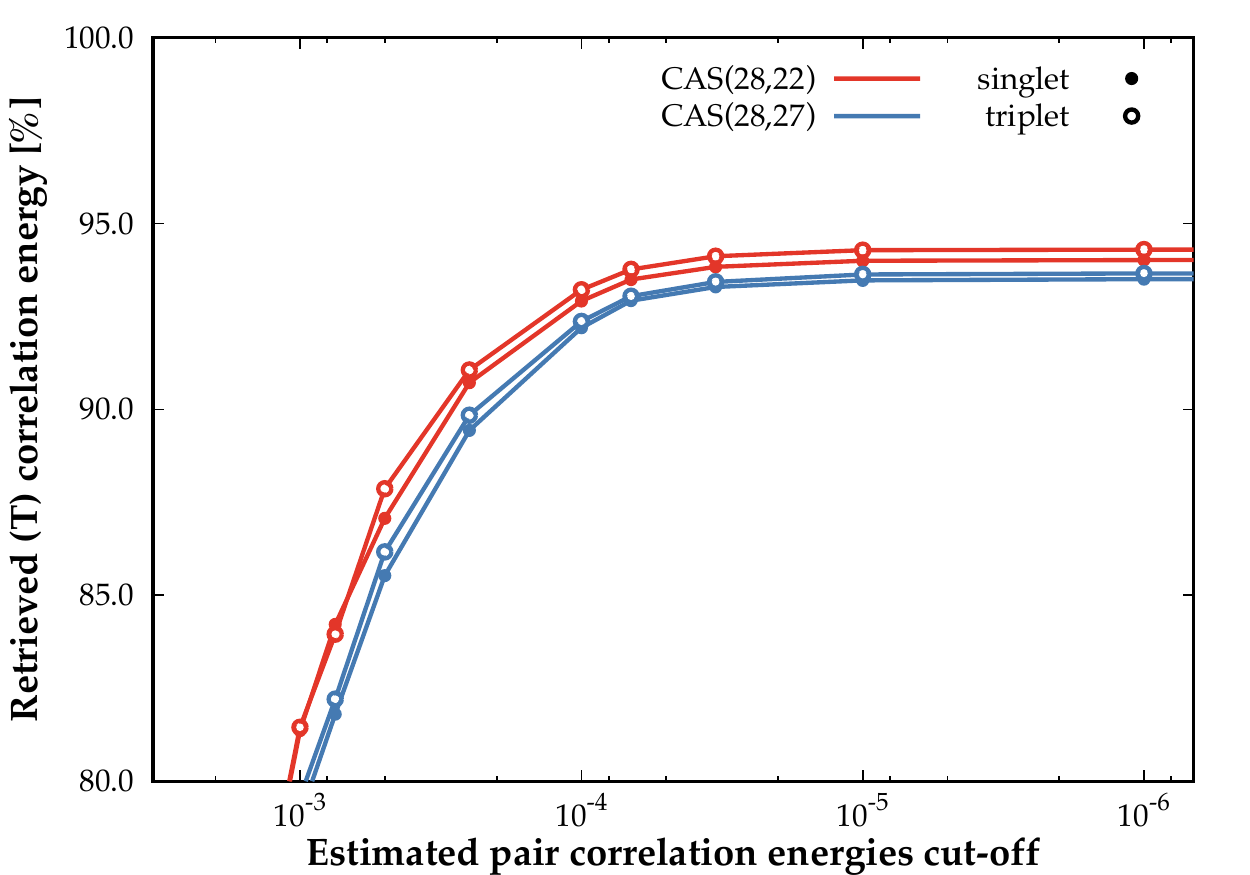}
  \hfill
  \includegraphics[width=0.48\textwidth]{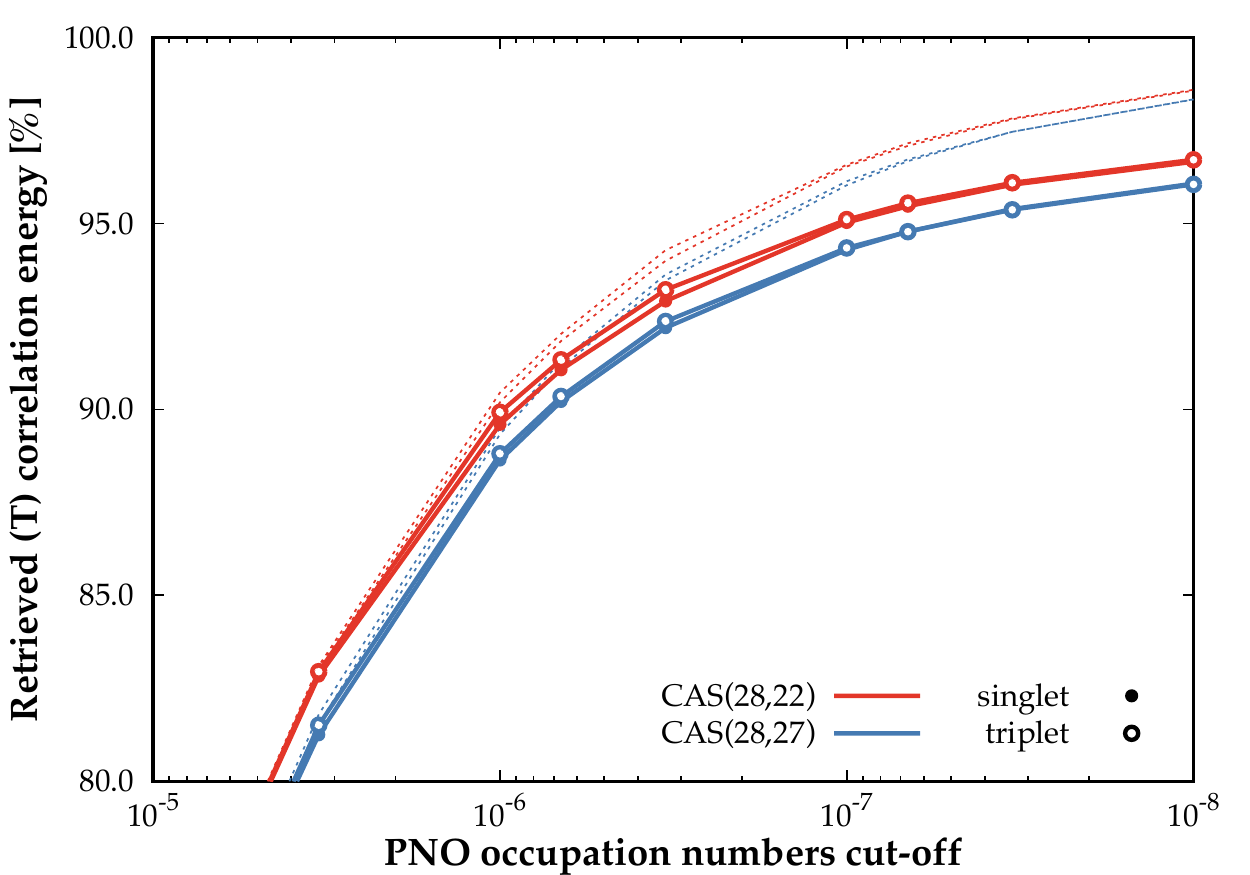}
  \caption{The percentage of perturbative triples correction correlation energy of oxo-Mn(Salen) in cc-pVDZ basis
  retrieved by DLPNO-TCCSD(T) with respect to canonical TCCSD(T) calculations as a function of cut-off for
  estimated pair correlation energies $T_\text{CutPairs}$ (left) and cut-off for PNO occupation numbers
  $T_\text{CutPNO}$ (right).
  The thin dashed lines in the right plot represent the values for $T_\text{CutPairs} = 10^{-5}$.}
  \label{fig:oxo_T}
\end{figure*}

Again, we first assessed the percentage of the recovered canonical TCCSD correlation energy with respect
to the cut-off parameter $T_\text{CutPairs}$.
The curve shown in the left plot of Figure \ref{fig:oxo_SD} follows the same trend as for TME
and quickly converges.
This means that the retrieved correlation energy is already converged by the value $T_\text{CutPairs}=10^{-5}$.
Even though the correlation energy is not yet stable for the default value $10^{-4}$, the difference
in accuracy between two spin states is minimal.
Since we are interested in the width of the singlet-triplet gap, it appears that even this setting
provides quite reasonable accuracy.

The right plot of Figure \ref{fig:oxo_SD} shows the dependence on the cut-off parameter $T_\text{CutPNO}$.
These curves smoothly near towards 100\% as the parameter tightens, the behavior as observed for conventional
DLPNO-CCSD.
However, the energies overshoot for the very conservative values of the parameter, which is caused
by overcompensation of the neglected pairs with the MP2 pair energy.
For this reason, we calculated the same dependence also with the tighter setting of
$T_\text{CutPairs}=10^{-5}$ (the thinner dotted lines in the plot).
Since for this value the correlation energies are basically converged with respect to $T_\text{CutPairs}$,
we can observe that the curves now smoothly converge towards 100\%.
The singlet-triplet gap seems reasonably accurate for the default value of $T_\text{CutPNO}$ and for more
conservative values the difference in accuracy completely disappears.

\begin{figure}
  \centering
  \includegraphics[width=0.48\textwidth]{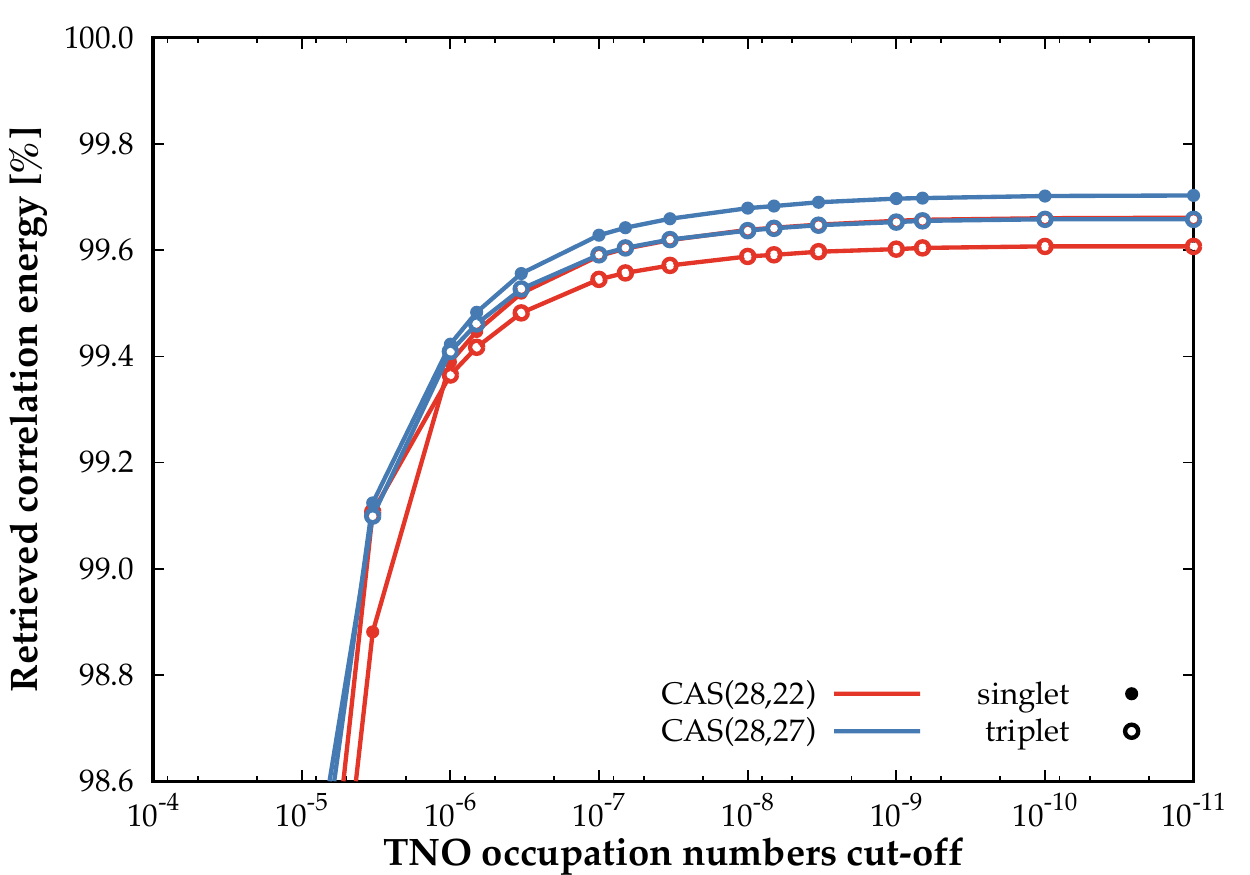}
  \caption{The percentage of correlation energy of oxo-Mn(Salen) in cc-pVDZ basis retrieved by DLPNO-TCCSD(T)
  with respect to canonical TCCSD(T) calculations as a function of cut-off for TNO occupation numbers
  $T_\text{CutTNO}$.}
  \label{fig:oxo_tno}
\end{figure}

We also examined the effect of these cut-offs on the perturbative triples correction recovered by DLPNO-TCCSD(T),
which are shown in Figure \ref{fig:oxo_T}.
At the default value of $T_\text{CutPairs}$ the energies seem to be almost converged and are fully converged
at the TightPNO setting, the actually relevant relative accuracy does not change.
In case of $T_\text{CutPNO}$, the curves seem to slowly converge towards 100\%, which is even more
apparent when tighter cut-off on pair energies is in place.
Note, that these values represent only the percentage of the canonical triples correction not the total
correlation energy and for this particular system, the triples correction amounts to less than 3\% of
the total correlation energy.
On top of that, we investigated the sensitivity of overall correlation energy on the $T_\text{CutTNO}$
with the conclusion that the default value is more than adequate, see Figure \ref{fig:oxo_tno}.

\begin{table}
  \caption{Energy differences in kcal/mol between DLPNO-TCCSD and DLPNO-CCSD(T) calculations with different
           settings of cut-off parameters and equivalent canonical calculations on oxo-Mn(Salen) in cc-pVDZ.}
  \def\arraystretch{1.2}
  \begin{tabular}{@{\extracolsep{6pt}}cccccc}
    &      & \multicolumn{2}{c}{CAS(28,22)} & \multicolumn{2}{c}{CAS(28,27)} \\ \cline{3-4} \cline{5-6}
                         &         & default      & TightPNO     & default      & TightPNO     \\ \hline
    \multirow{3}{*}{SD}  & S       & $2.52$ & $3.45$ & $3.60$ & $4.30$ \\
                         & T       & $1.92$ & $3.01$ & $3.06$ & $3.97$ \\
         & $\Delta E_\text{S--T}$  & $0.60$ & $0.44$ & $0.54$ & $0.33$ \\ \hline
    \multirow{3}{*}{SD(T)}  & S    & $6.38$ & $5.30$ & $7.33$ & $6.19$ \\
                            & T    & $5.54$ & $4.79$ & $6.67$ & $5.80$ \\
         & $\Delta E_\text{S--T}$  & $0.84$ & $0.51$ & $0.66$ & $0.39$
  \end{tabular}
  \label{tab:oxo_diffs}
\end{table}

\begin{figure}
  \centering
  \includegraphics[width=0.48\textwidth]{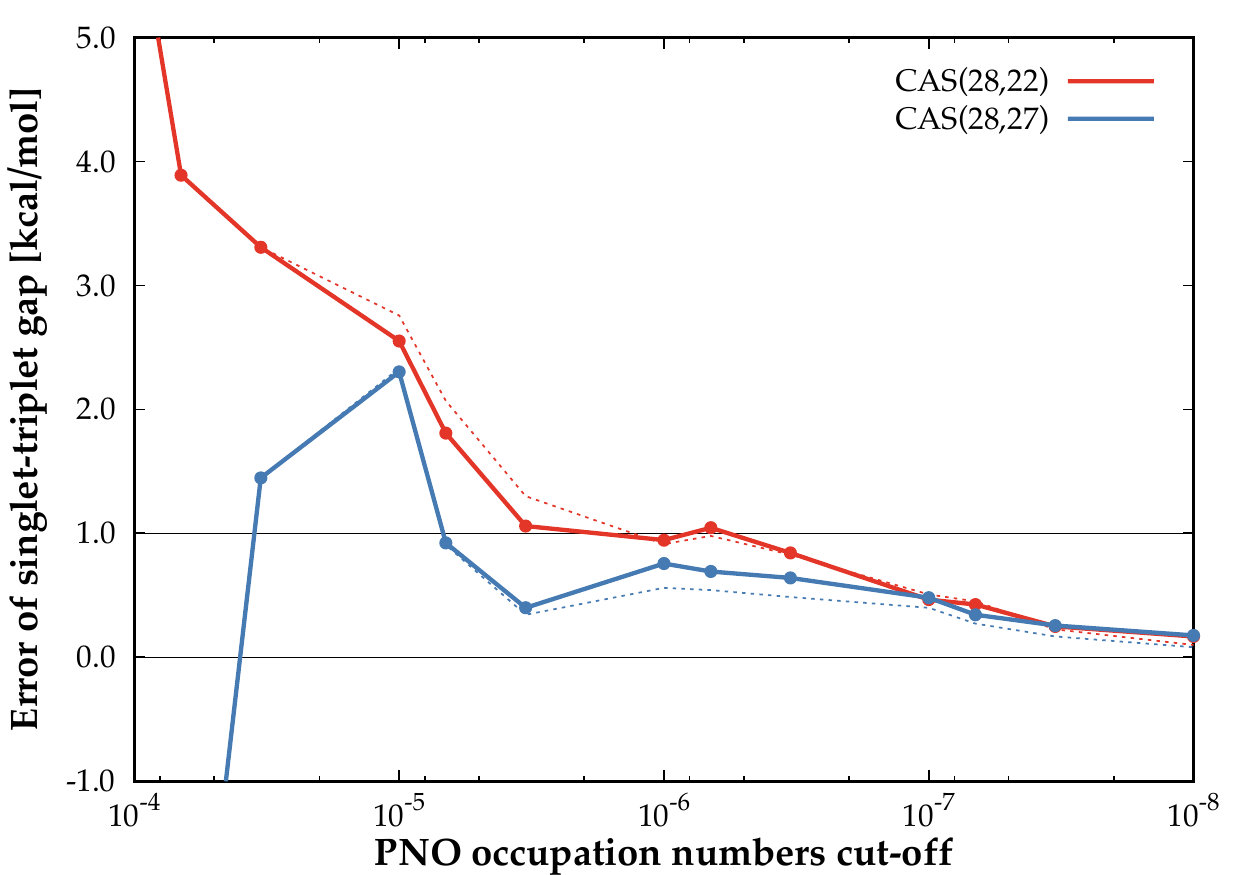}
  \caption{The error of DLPNO-TCCSD(T) in singlet-triplet gap of oxo-Mn(Salen) in cc-pVDZ basis with respect to
  canonical TCCSD(T) calculations as a function of cut-off for PNO occupation numbers $T_\text{CutPNO}$.
  The thin dashed lines represent the values for $T_\text{CutPairs} = 10^{-5}$.}
  \label{fig:oxo_gap}
\end{figure}

Table \ref{tab:oxo_diffs} contains the differences between DLPNO and canonical energies for
the singlet-triplet gap.
Although the errors in energies for singlet or triplet state alone are substantial (especially for the larger CAS),
the errors for the gap are well within the chemical accuracy of 1 kcal/mol.
When the perturbative triples correction is invoked, tighter cut-offs are preferable.
The same can be observed in Figure \ref{fig:oxo_gap}, where the chemical accuracy is achieved even with
looser than default settings.
Also the behavior of the DLPNO approximation seems quite stable with respect to the size of the active space.

When we compare these results to the previous LPNO-CCSD calculations \cite{Antalik2019}, the DLPNO-TCCSD
has smaller error in absolute energies, but larger in gaps.
This can be attributed to the rather fortunate cancellation of errors in the LPNO case, but otherwise
the DLPNO implementation is more reliable due to smaller errors in absolute energies and smooth convergence
towards the canonical correlation energies.

Moreover, we performed calculations with up to cc-pVQZ basis set.
We found the results to be consistent with the previous LPNO-TCCSD calculations and NEVPT2(28,22), with
perturbative triples correction being responsible only for a minor shift in energy, see Table \ref{tab:oxo_gap}.
Even though the calculation in cc-pVQZ, which amounts to 1178 basis functions, is perfectly within the
possibilities of the LPNO approach, it took 60\% longer to finish under the same conditions (8 cores, 30GB
of memory per core) compared to the DLPNO calculation.

\begin{table}
	\def\arraystretch{1.2}
	\setlength{\tabcolsep}{1em}
	\caption{The singlet and triplet energies of oxo-Mn(salen) in cc-pVXZ basis sets. the difference
  $E(^1\text{A})-E(^3\text{A})$  in kcal/mol. Results for different active spaces
  and in various basis sets.}
	\begin{tabular}{lrrr}
    &  \multicolumn{1}{c}{\ph{--}DZ} &  \multicolumn{1}{c}{\ph{--}TZ} &  \multicolumn{1}{c}{\ph{--}QZ} \\ \hline
		LPNO-TCCSD(28,22)\cite{Antalik2019}  &  $6.2$  &  $6.3$  &  $6.3$  \\
		DLPNO-TCCSD(28,22)                    &  $6.9$  &  $7.3$  &  $6.9$  \\
		DLPNO-TCCSD(T)(28,22)                 &  $5.9$  &  $6.3$  &  $5.8$  \\ \hline
		LPNO-TCCSD(28,27)\cite{Antalik2019}  &  $3.7$  &  $3.1$  &  $2.9$  \\
		DLPNO-TCCSD(28,27)                    &  $4.1$  &  $4.0$  &  $3.1$  \\
		DLPNO-TCCSD(T)(28,27)                 &  $4.0$  &  $5.0$  &  $2.9$  \\ \hline
    NEVPT2(28,22)\cite{Sharma2017}       & $-7.4$  &  $1.6$  &  $2.4$  
	\end{tabular}
	\label{tab:oxo_gap}
\end{table}

\begin{figure*}
  \centering
  \includegraphics[width=0.48\textwidth]{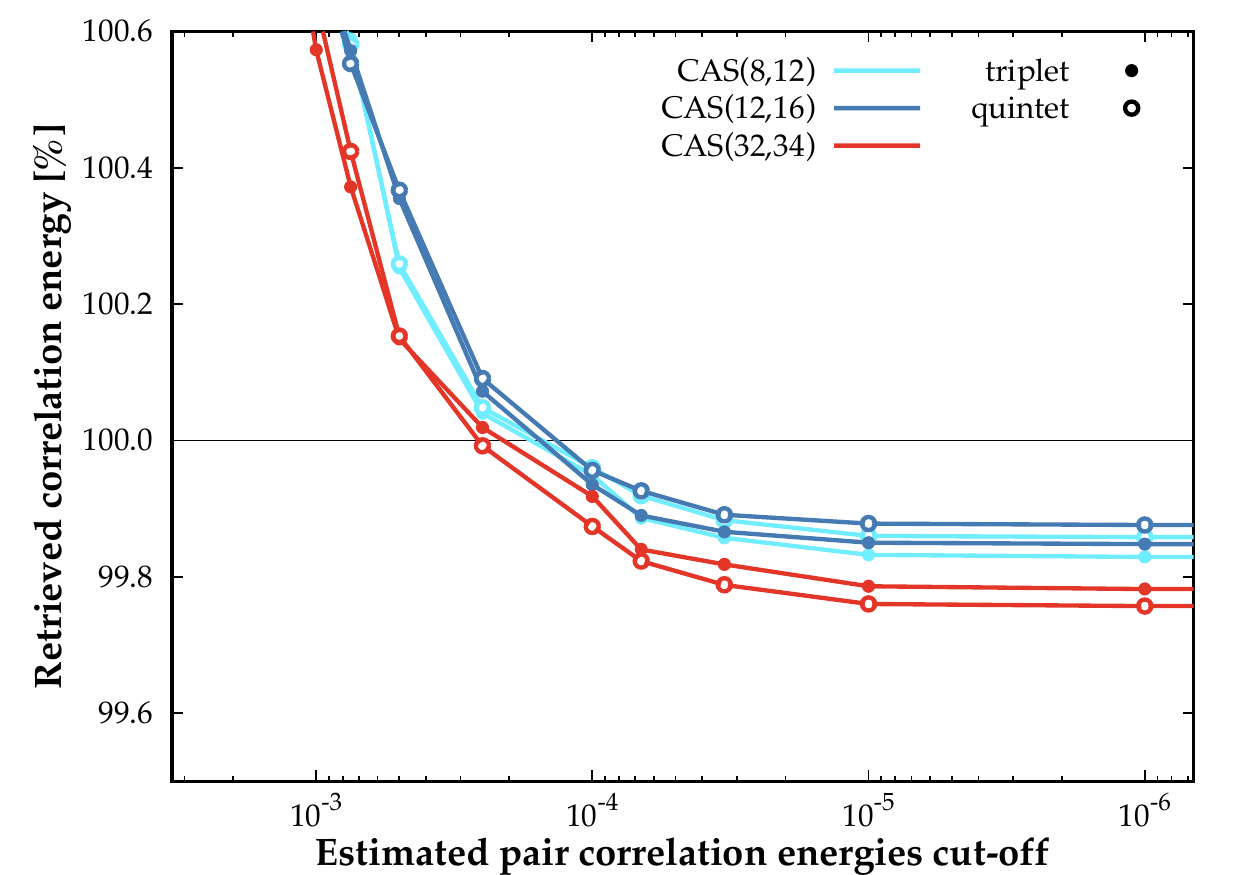}
  \hfill
  \includegraphics[width=0.48\textwidth]{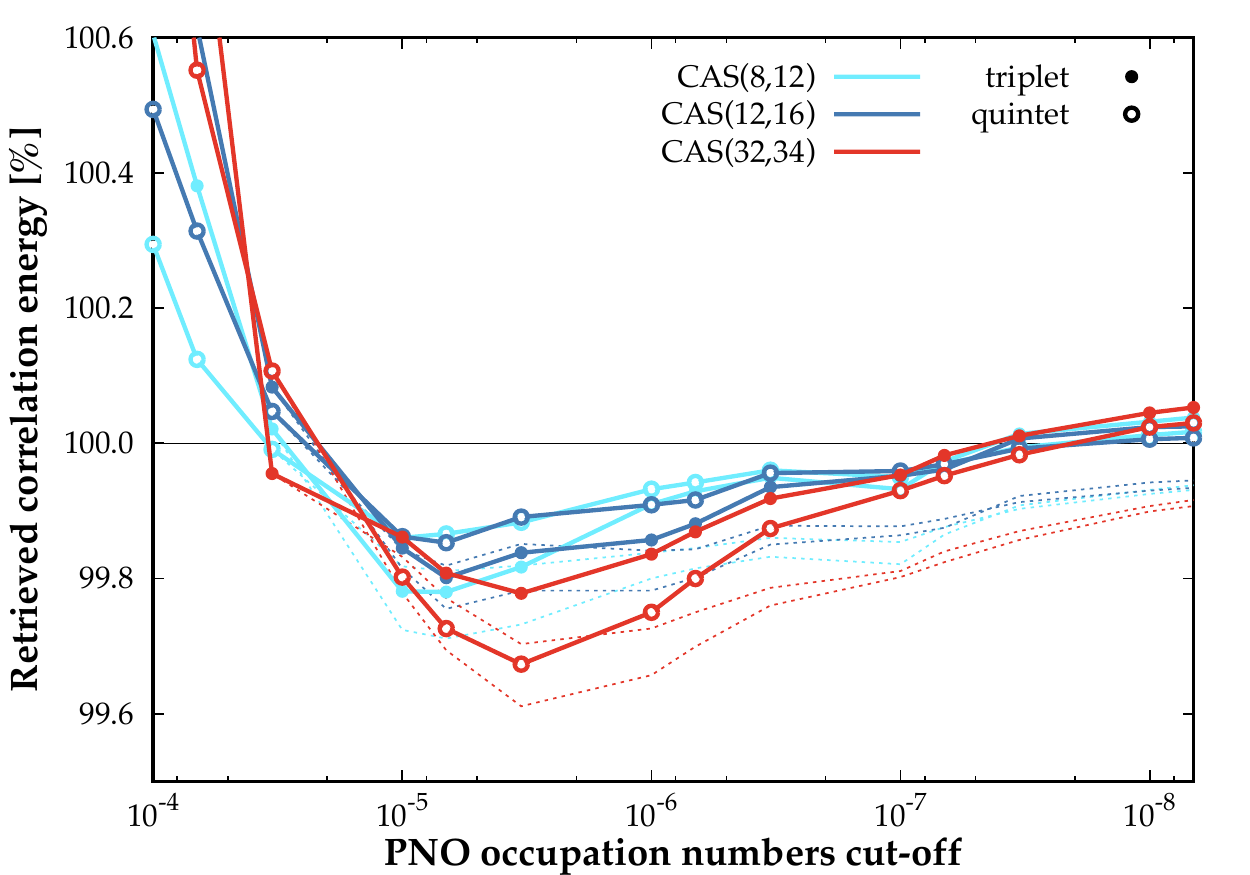}
  \caption{The percentage of correlation energy of FeP in def2-SVP basis retrieved by DLPNO-TCCSD
  with respect to canonical TCCSD calculations as a function of cut-off for estimated pair correlation
  energies $T_\text{CutPairs}$ (left) and cut-off for PNO occupation numbers $T_\text{CutPNO}$ (right).
  The thin dashed lines in the right plot represent the values for $T_\text{CutPairs} = 10^{-5}$.}
  \label{fig:fep_SD}
\end{figure*}

\begin{figure*}
  \centering
  \includegraphics[width=0.48\textwidth]{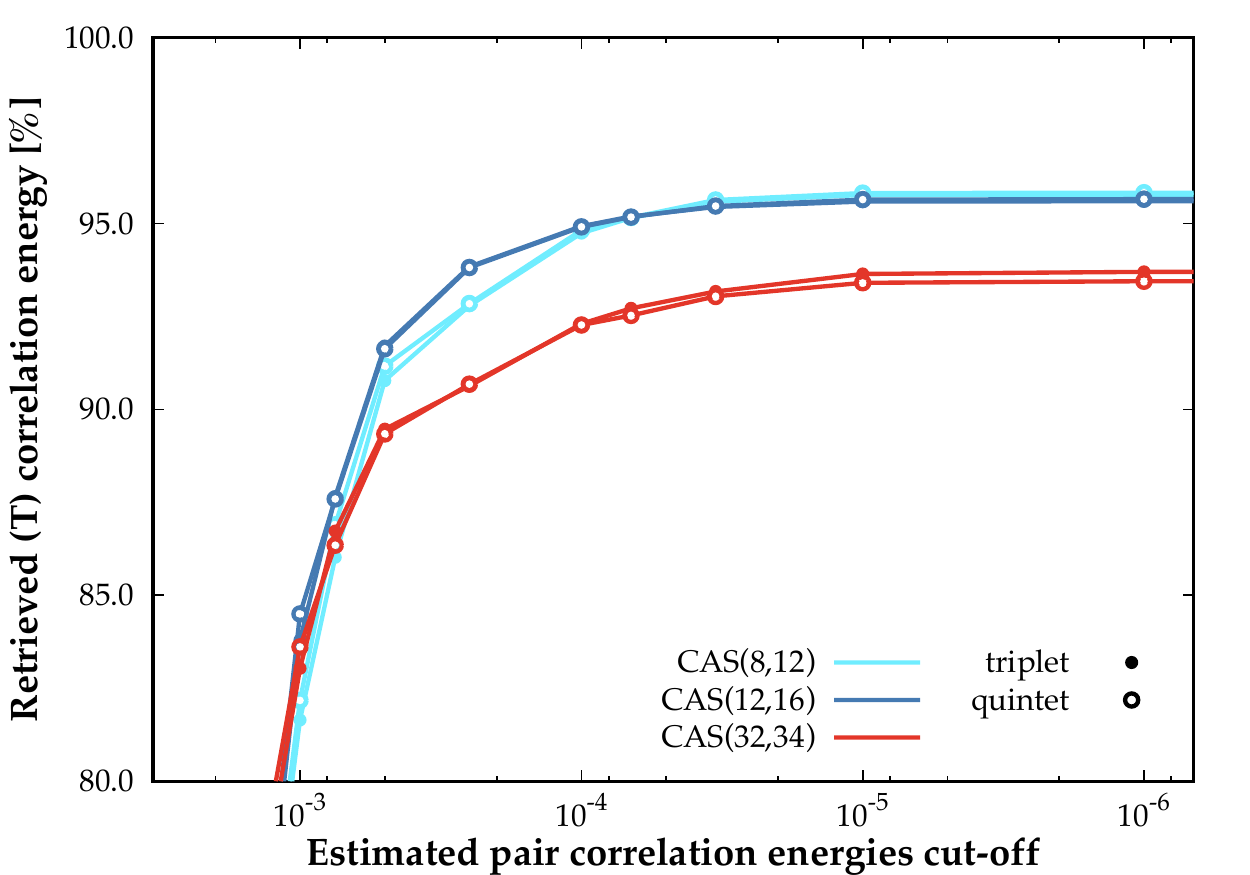}
  \hfill
  \includegraphics[width=0.48\textwidth]{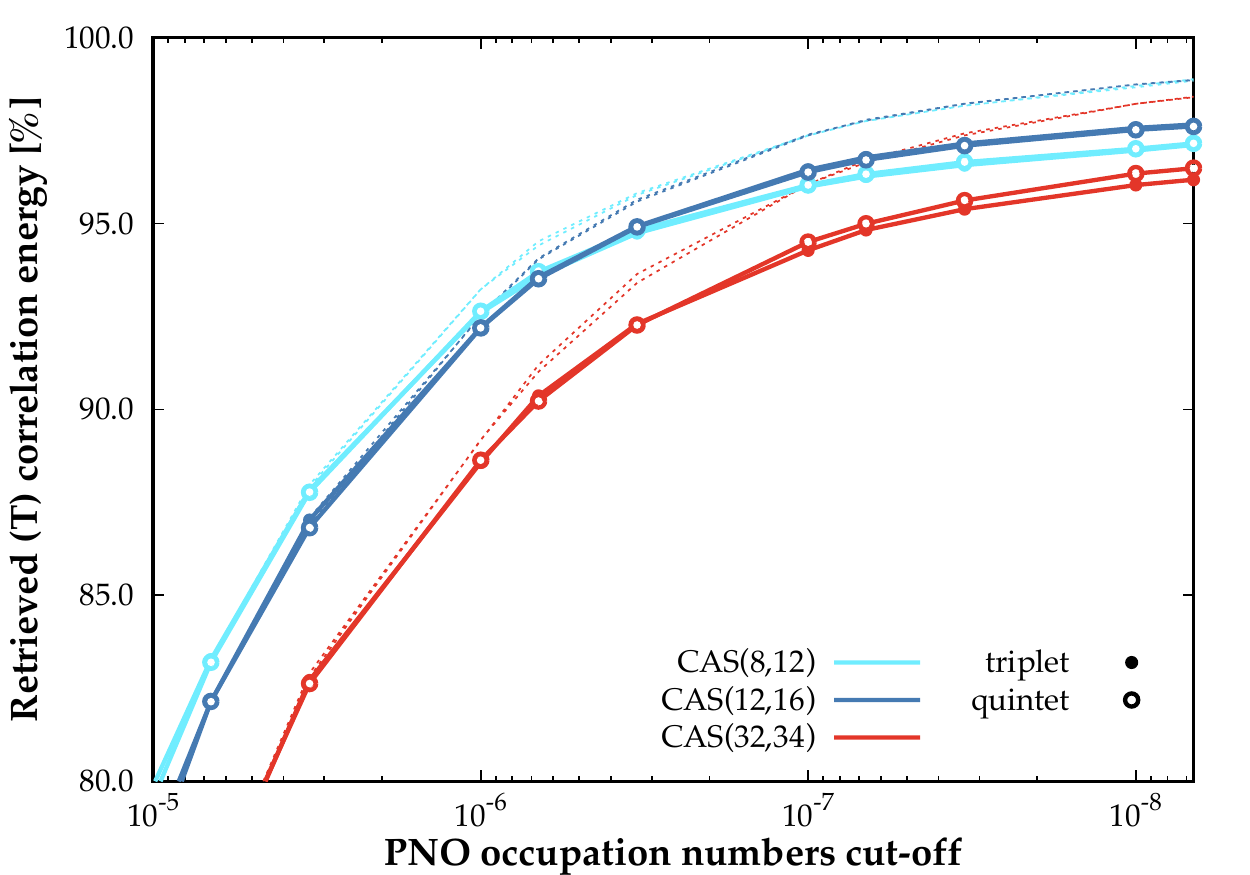}
  \caption{The percentage of perturbative triples correction correlation energy of FeP in def2-SVP basis
  retrieved by DLPNO-TCCSD(T) with respect to canonical TCCSD(T) calculations as a function of cut-off for
  estimated pair correlation energies $T_\text{CutPairs}$ (left) and cut-off for PNO occupation numbers
  $T_\text{CutPNO}$ (right).
  The thin dashed lines in the right plot represent the values for $T_\text{CutPairs} = 10^{-5}$.}
  \label{fig:fep_T}
\end{figure*}
 
\subsection{Iron(II) Porphyrin Model}
Fe(II) porphyrin derivatives play important roles in reactions related to material science and biological
processes due to their closely lying spin states.
The chosen model has previously been a subject of several large scale CASSCF studies\cite{LiManni2018,LiManni2019}
and we therefore consider it to be an interesting system to test the efficiency of the method for several active
spaces of different size.

The left plot in Figure \ref{fig:fep_SD} shows the dependence on the $T_\text{CutPairs}$ cut-off parameter.
As for oxo-Mn(Salen), the curves converge smoothly and at $10^{-5}$ they are practically converged. 
The accuracy is consistent for triplet and quintet spin states and it differs maximally by 0.05\% of
the retrieved canonical correlation energy.
This is true regardless of the size of the active space, although the overall accuracy is slightly lower
for the largest CAS.

The right plot in Figure \ref{fig:fep_SD} shows the dependence on the $T_\text{CutPNO}$ cut-off parameter.
For very loose values, the discrepancy in accuracy for different spin states is apparent, especially for the
larger CAS.
However, it disappears with default and TightPNO settings of the cut-off.
The method once again overestimates the correlation energy for the default value of $T_\text{CutPairs}$, but
for tighter value $T_\text{CutPairs}=10^{-5}$ it converges smoothly towards 100\%.

\begin{figure}
	\centering
  \includegraphics[width=0.48\textwidth]{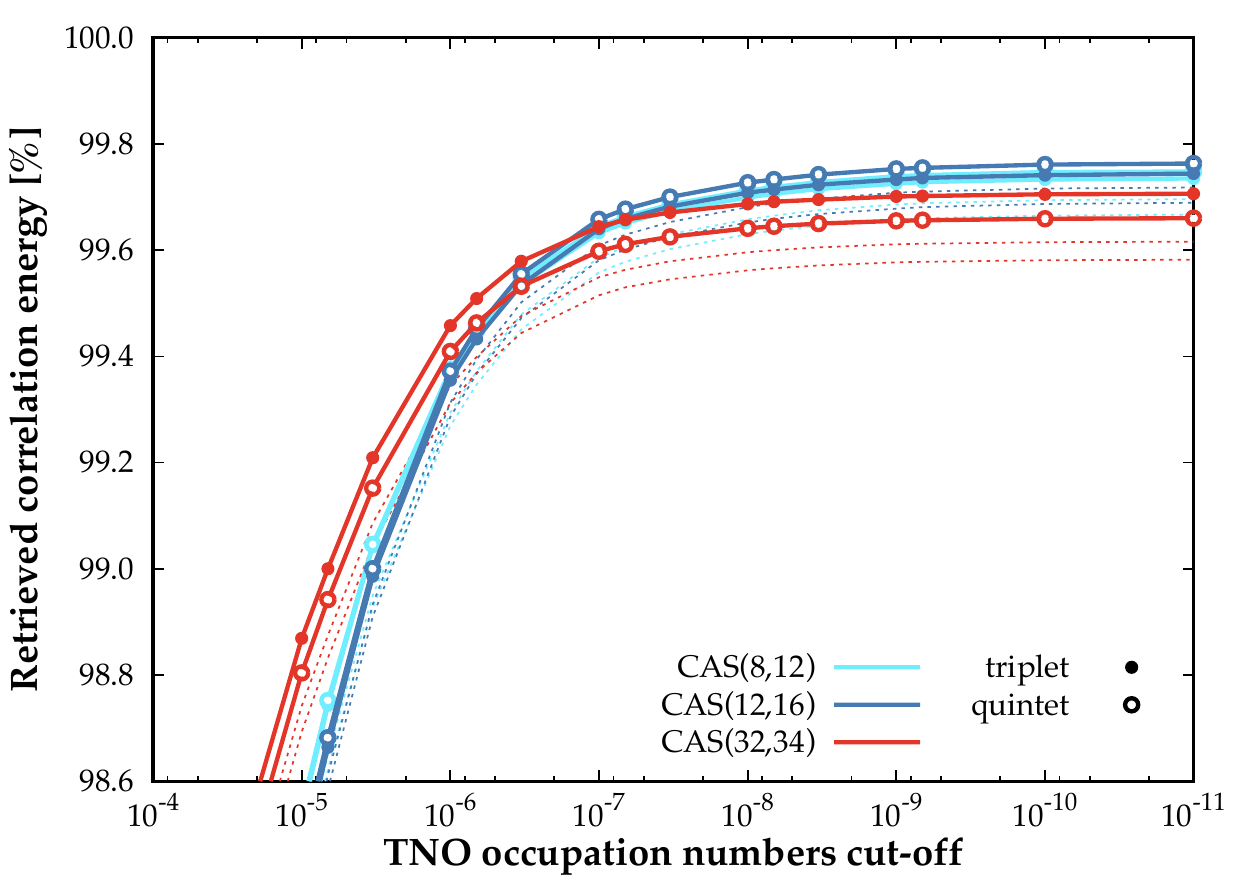}
  \caption{The percentage of correlation energy of FeP in cc-pVDZ basis retrieved by DLPNO-TCCSD(T)
  with respect to canonical TCCSD(T) calculations as a function of cut-off for TNO occupation numbers
  $T_\text{CutTNO}$.
  The thin dashed lines represent the values for $T_\text{CutPairs} = 10^{-5}$.}
  \label{fig:fep_tno}
\end{figure}

\begin{table*}
	\def\arraystretch{1.2}
	\setlength{\tabcolsep}{0.5em}
	\caption{Energy differences in kcal/mol between DLPNO-TCCSD and DLPNO-CCSD(T) calculations with different
           settings of cut-off parameters and equivalent canonical calculations on Fe(II)-porphyrin model
           in def2-SVP.}
   \begin{tabular}{@{\extracolsep{6pt}}cccccccc}
    &      & \multicolumn{2}{c}{CAS(8,12)} & \multicolumn{2}{c}{CAS(12,16)} & \multicolumn{2}{c}{CAS(32,34)} \\
             \cline{3-4}                     \cline{5-6}                      \cline{7-8}
                           &       & default&TightPNO&default &TightPNO& default            & TightPNO     \\ \hline
     \multirow{3}{*}{SD}   & T     & $0.85$ & $3.01$ & $1.10$ & $2.27$ & $\ph{-}1.42\ph{-}$ & $\ph{-}3.28\ph{-}$ \\
                           & Q     & $0.69$ & $2.41$ & $0.73$ & $2.04$ & $\ph{-}2.16\ph{-}$ & $\ph{-}3.40\ph{-}$ \\
         & $\Delta E_\text{T--Q}$  & $0.16$ & $0.60$ & $0.37$ & $0.23$ &      $-0.74\ph{-}$ & $-0.12\ph{-}$ \\ \hline
    \multirow{3}{*}{SD(T)} & T     & $4.87$ & $5.03$ & $4.77$ & $4.20$ & $\ph{-}5.49\ph{-}$ & $\ph{-}5.43\ph{-}$ \\
                           & Q     & $4.58$ & $4.38$ & $4.34$ & $3.90$ & $\ph{-}6.19\ph{-}$ & $\ph{-}5.52\ph{-}$ \\
         & $\Delta E_\text{T--Q}$  & $0.29$ & $0.65$ & $0.43$ & $0.30$ &      $-0.70\ph{-}$ &      $-0.09\ph{-}$
  \end{tabular}
	\label{tab:fep_gap}
\end{table*} 

\begin{figure}
  \centering
  \includegraphics[width=0.48\textwidth]{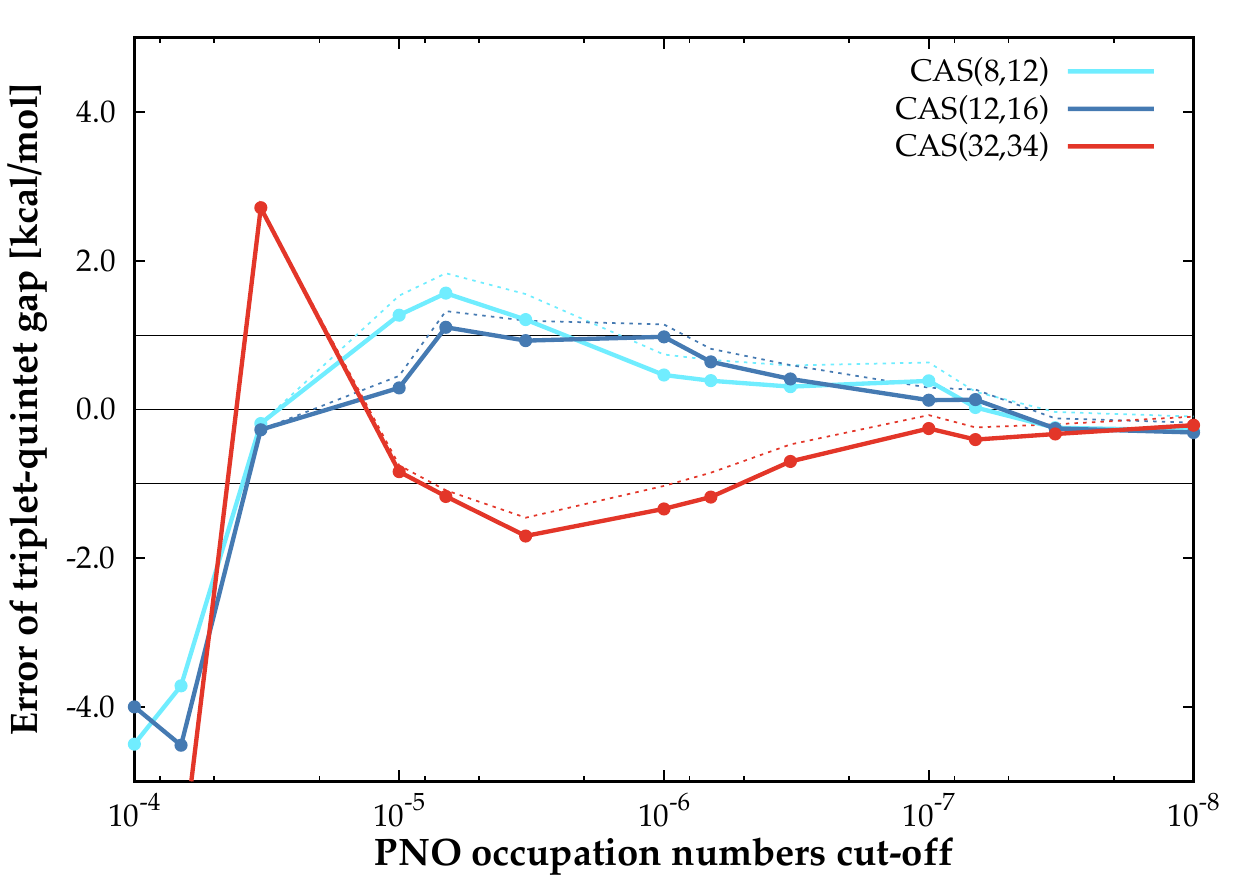}
  \caption{The error of DLPNO-TCCSD(T) in singlet-triplet gap of FeP in cc-pVDZ basis with respect to
  canonical TCCSD(T) calculations as a function of cut-off for PNO occupation numbers $T_\text{CutPNO}$.}
  \label{fig:fep_gap}
\end{figure}

Regarding the triples correction, the convergence behavior is similar as for oxo-Mn(Salen).
The percentages of retrieved triples energy with respect to both $T_\text{CutPairs}$ and $T_\text{CutPNO}$
are shown in Figure \ref{fig:fep_T}.
The DLPNO approximation for this correction is again the most sensitive to changes in $T_\text{CutPNO}$ parameter,
but shows clear convergence towards 100\%.
Dependence of total DLPNO correlation energy on $T_\text{CutTNO}$ is shown in Figure \ref{fig:fep_tno}.

Finally, we assess the accuracy of the triplet-quintet gap, which is the actual value of interest.
Table \ref{tab:fep_gap} lists the energy differences between TCCSD, TCCSD(T) and their DLPNO versions.
At the CCSD level, the errors for absolute energies grow with growing CAS.
These errors are larger for the TightPNO settings, which is understandable given the method overshoots for
looser $T_\text{CutPairs}$ thresholds.
However, the tighter cut-offs significantly improve the accuracy in energy of the triplet-quintet gap,
with the exception of the smallest active space.
Although the decrease in accuracy of the absolute energies is observed, once the perturbative triples correction is included, the errors in gaps are basically the same as in the CCSD case.
Moreover, the dependence of the DLPNO error on $T_\text{CutPNO}$ is presented in Figure \ref{fig:fep_gap}.
Even for the default cut-offs, the error is well under 1 kcal/mol and is even smaller for the TightPNO
settings, with the previously discussed exception of CAS(8,12).
Interestingly, the accuracy of the quintet state is consistently lower than for the triplet state for
the largest CAS, while it is the opposite for the smaller active spaces.

\section{Conclusions}
\label{sec_conclusions}

We introduced a new version of DMRG-TCCSD method, which employs the domain-based local pair natural orbital approach.
The method was implemented in ORCA and should be available in the next release, while the Budapest DMRG code
is used for the DMRG part.

We performed several accuracy assessments of the method employing three systems, two of which have been previously
studied by the canonical and LPNO versions of the TCCSD method.

For tetramethyleneethane, we were able to retrieve over 99.9\% of the canonical correlation energy, while
using the default settings of cut-off parameters, with negligible non-parallelity error with respect to its
dihedral rotation.
Also, the drawbacks present in the LPNO version were eliminated.

For oxo-Mn(Salen), the amount of retrieved correlation was dependent on the size of the used active space,
ranging from 99.8\% for the larger CAS(28,27) to 99.9\% for smaller CAS(28,22), which is an improvement
about 0.2\% with respect to LPNO-TCCSD.
Using the default settings resulted in singlet-triplet gap being off by 0.5-0.6 kcal/mol and with tighter cut-offs
only 0.3-0.4 kcal/mol compared to canonical calculation.
These results are slightly worse than those of LPNO-TCCSD, which we believe is a consequence of a fortunate
cancellation of errors in LPNO-TCCSD, since DLPNO-TCCSD are consistently better for the remaining two systems.
Moreover, new implementation offers significantly better timing than the LPNO one.

For Fe(II)-porphyrin, the dependence on cut-off parameters was very similar as for oxo-Mn(Salen).
With the default settings, we were able to retrieve more than 99.9\% canonical correlation
energy.
Irrespective of the active space size, the method determined the triplet-quintet gap with error safely within 
the chemical accuracy of 1kcal/mol even with the default settings.

To summarize, we believe that DLPNO-DMRG-TCCSD(T) is a possible approach to treat systems with a moderate
multireference character. 
In the near future, we would like to implement the DLPNO version of multireference TCCSD,
which we hope to further enhance capabilities of this method.
\vfill

\section*{Acknowledgment}

We would like to thank Prof. Frank Neese for providing us with access to ORCA source code,
as well as for helpful discussions, Dr. Frank Wennmohs for technical assistance with the ORCA code
and Dr. Ond\v{r}ej Demel for helpful discussions.

This work has been supported by
the Czech Science Foundation (Grants No. 18-24563S, 16-12052S and 18-18940Y),
the Hungarian National Research, Development and Innovation Office (NKFIH) through (Grant No. K120569),
the Hungarian Quantum Technology National Excellence Program (Project No. 2017-1.2.1-NKP-2017-00001),
by the Pacific Northwest National Laboratory (Contract No. 462268),
and
by the Czech Ministry of Education, Youth and Sports (Project No. LTAUSA17033 and the Large Infrastructures
for Research, Experimental Development and Innovations project „IT4Innovations National Supercomputing
Center – LM2015070“.
{\"O}. Legeza acknowledges financial support from the Alexander von Humboldt foundation.
Mutual visits with members of the Hungarian group have been partly supported by the Hungarian-Czech
Joint Research Project MTA/19/04.

\bibliography{new}

\providecommand{\latin}[1]{#1}
\makeatletter
\providecommand{\doi}
  {\begingroup\let\do\@makeother\dospecials
  \catcode`\{=1 \catcode`\}=2 \doi@aux}
\providecommand{\doi@aux}[1]{\endgroup\texttt{#1}}
\makeatother
\providecommand*\mcitethebibliography{\thebibliography}
\csname @ifundefined\endcsname{endmcitethebibliography}
  {\let\endmcitethebibliography\endthebibliography}{}
\begin{mcitethebibliography}{150}
\providecommand*\natexlab[1]{#1}
\providecommand*\mciteSetBstSublistMode[1]{}
\providecommand*\mciteSetBstMaxWidthForm[2]{}
\providecommand*\mciteBstWouldAddEndPuncttrue
  {\def\EndOfBibitem{\unskip.}}
\providecommand*\mciteBstWouldAddEndPunctfalse
  {\let\EndOfBibitem\relax}
\providecommand*\mciteSetBstMidEndSepPunct[3]{}
\providecommand*\mciteSetBstSublistLabelBeginEnd[3]{}
\providecommand*\EndOfBibitem{}
\mciteSetBstSublistMode{f}
\mciteSetBstMaxWidthForm{subitem}{(\alph{mcitesubitemcount})}
\mciteSetBstSublistLabelBeginEnd
  {\mcitemaxwidthsubitemform\space}
  {\relax}
  {\relax}

\bibitem[{\v{C}}{\'{\i}}{\v{z}}ek(1966)]{Cizek1966}
{\v{C}}{\'{\i}}{\v{z}}ek,~J. On the Correlation Problem in Atomic and Molecular
  Systems. Calculation of Wavefunction Components in Ursell-Type Expansion
  Using Quantum-Field Theoretical Methods. \emph{J. Chem. Phys.} \textbf{1966},
  \emph{45}, 4256--4266\relax
\mciteBstWouldAddEndPuncttrue
\mciteSetBstMidEndSepPunct{\mcitedefaultmidpunct}
{\mcitedefaultendpunct}{\mcitedefaultseppunct}\relax
\EndOfBibitem
\bibitem[Gauss(1998)]{Gauss1998_encyclop}
Gauss,~J. In \emph{The Encyclopedia of Computational Chemistry};
  v.~R.~Schleyer,~P., Allinger,~N.~L., Clark,~T., Gasteiger,~J.,
  Kollman,~P.~A., {Schaefer~III},~H.~F., Scheiner,~P.~R., Eds.; Wiley:
  Chichester, 1998; pp 615--636\relax
\mciteBstWouldAddEndPuncttrue
\mciteSetBstMidEndSepPunct{\mcitedefaultmidpunct}
{\mcitedefaultendpunct}{\mcitedefaultseppunct}\relax
\EndOfBibitem
\bibitem[Raghavachari \latin{et~al.}(1989)Raghavachari, Trucks, Pople, and
  Head-Gordon]{Raghavachari1989}
Raghavachari,~K.; Trucks,~G.~W.; Pople,~J.~A.; Head-Gordon,~M. A fifth-order
  perturbation comparison of electron correlation theories. \emph{Chem. Phys.
  Lett.} \textbf{1989}, \emph{157}, 479--483\relax
\mciteBstWouldAddEndPuncttrue
\mciteSetBstMidEndSepPunct{\mcitedefaultmidpunct}
{\mcitedefaultendpunct}{\mcitedefaultseppunct}\relax
\EndOfBibitem
\bibitem[Bartlett and Musia{\l}(2007)Bartlett, and Musia{\l}]{Bartlett2007}
Bartlett,~R.~J.; Musia{\l},~M. Coupled-cluster theory in quantum chemistry.
  \emph{Rev. Mod. Phys.} \textbf{2007}, \emph{79}, 291--352\relax
\mciteBstWouldAddEndPuncttrue
\mciteSetBstMidEndSepPunct{\mcitedefaultmidpunct}
{\mcitedefaultendpunct}{\mcitedefaultseppunct}\relax
\EndOfBibitem
\bibitem[Tew \latin{et~al.}(2010)Tew, H{\"{a}}ttig, Bachorz, and
  Klopper]{Tew2010}
Tew,~D.~P.; H{\"{a}}ttig,~C.; Bachorz,~R.~A.; Klopper,~W. In \emph{Recent
  Progress in Coupled Cluster Methods}; \v{C}\'arsky,~P., Paldus,~J.,
  Pittner,~J., Eds.; Springer Science: New York, 2010; p 535\relax
\mciteBstWouldAddEndPuncttrue
\mciteSetBstMidEndSepPunct{\mcitedefaultmidpunct}
{\mcitedefaultendpunct}{\mcitedefaultseppunct}\relax
\EndOfBibitem
\bibitem[Lyakh \latin{et~al.}(2012)Lyakh, Musia{\l}, Lotrich, and
  Bartlett]{Lyakh2012}
Lyakh,~D.~I.; Musia{\l},~M.; Lotrich,~V.~F.; Bartlett,~R.~J. Multireference
  Nature of Chemistry: The Coupled-Cluster View. \emph{Chem. Rev.}
  \textbf{2012}, \emph{112}, 182--243\relax
\mciteBstWouldAddEndPuncttrue
\mciteSetBstMidEndSepPunct{\mcitedefaultmidpunct}
{\mcitedefaultendpunct}{\mcitedefaultseppunct}\relax
\EndOfBibitem
\bibitem[Li and Paldus(1997)Li, and Paldus]{Li1997a}
Li,~X.; Paldus,~J. Reduced multireference {CCSD} method: An effective approach
  to quasidegenerate states. \emph{J. Chem. Phys.} \textbf{1997}, \emph{107},
  6257--6269\relax
\mciteBstWouldAddEndPuncttrue
\mciteSetBstMidEndSepPunct{\mcitedefaultmidpunct}
{\mcitedefaultendpunct}{\mcitedefaultseppunct}\relax
\EndOfBibitem
\bibitem[Li(2001)]{Li2001_theochem}
Li,~X. Benchmark study of potential energies and vibrational levels using the
  reduced multireference coupled cluster method. The {HF} molecule. \emph{J.
  Mol. Struct.: THEOCHEM} \textbf{2001}, \emph{547}, 69--81\relax
\mciteBstWouldAddEndPuncttrue
\mciteSetBstMidEndSepPunct{\mcitedefaultmidpunct}
{\mcitedefaultendpunct}{\mcitedefaultseppunct}\relax
\EndOfBibitem
\bibitem[Li and Paldus(2001)Li, and Paldus]{Li2001_jcp_a}
Li,~X.; Paldus,~J. Energy versus amplitude corrected coupled-cluster
  approaches. {II}. Breaking the triple bond. \emph{J. Chem. Phys.}
  \textbf{2001}, \emph{115}, 5774--5783\relax
\mciteBstWouldAddEndPuncttrue
\mciteSetBstMidEndSepPunct{\mcitedefaultmidpunct}
{\mcitedefaultendpunct}{\mcitedefaultseppunct}\relax
\EndOfBibitem
\bibitem[Li and Paldus(2001)Li, and Paldus]{Li2001_jcp_b}
Li,~X.; Paldus,~J. Energy versus amplitude corrected coupled-cluster
  approaches. I. \emph{J. Chem. Phys.} \textbf{2001}, \emph{115},
  5759--5773\relax
\mciteBstWouldAddEndPuncttrue
\mciteSetBstMidEndSepPunct{\mcitedefaultmidpunct}
{\mcitedefaultendpunct}{\mcitedefaultseppunct}\relax
\EndOfBibitem
\bibitem[Paldus and Planelles(1994)Paldus, and Planelles]{Paldus1994}
Paldus,~J.; Planelles,~J. Valence bond corrected single reference coupled
  cluster approach. \emph{Theor. Chim. Acta} \textbf{1994}, \emph{89},
  13--31\relax
\mciteBstWouldAddEndPuncttrue
\mciteSetBstMidEndSepPunct{\mcitedefaultmidpunct}
{\mcitedefaultendpunct}{\mcitedefaultseppunct}\relax
\EndOfBibitem
\bibitem[Piecuch \latin{et~al.}(1996)Piecuch, Tobol/a, and Paldus]{Piecuch1996}
Piecuch,~P.; Tobol/a,~R.; Paldus,~J. Approximate account of connected quadruply
  excited clusters in single-reference coupled-cluster theory via cluster
  analysis of the projected unrestricted Hartree-Fock wave function.
  \emph{Phys. Rev. A} \textbf{1996}, \emph{54}, 1210--1241\relax
\mciteBstWouldAddEndPuncttrue
\mciteSetBstMidEndSepPunct{\mcitedefaultmidpunct}
{\mcitedefaultendpunct}{\mcitedefaultseppunct}\relax
\EndOfBibitem
\bibitem[Li \latin{et~al.}(1997)Li, Peris, Planelles, Rajadall, and
  Paldus]{Li1997b}
Li,~X.; Peris,~G.; Planelles,~J.; Rajadall,~F.; Paldus,~J. Externally corrected
  singles and doubles coupled cluster methods for open-shell systems. \emph{J.
  Chem. Phys.} \textbf{1997}, \emph{107}, 90--98\relax
\mciteBstWouldAddEndPuncttrue
\mciteSetBstMidEndSepPunct{\mcitedefaultmidpunct}
{\mcitedefaultendpunct}{\mcitedefaultseppunct}\relax
\EndOfBibitem
\bibitem[Kinoshita \latin{et~al.}(2005)Kinoshita, Hino, and
  Bartlett]{Kinoshita2005}
Kinoshita,~T.; Hino,~O.; Bartlett,~R.~J. Coupled-cluster method tailored by
  configuration interaction. \emph{J. Chem. Phys.} \textbf{2005}, \emph{123},
  074106\relax
\mciteBstWouldAddEndPuncttrue
\mciteSetBstMidEndSepPunct{\mcitedefaultmidpunct}
{\mcitedefaultendpunct}{\mcitedefaultseppunct}\relax
\EndOfBibitem
\bibitem[Lyakh \latin{et~al.}(2011)Lyakh, Lotrich, and Bartlett]{Lyakh2011}
Lyakh,~D.~I.; Lotrich,~V.~F.; Bartlett,~R.~J. The `tailored' {CCSD}(T)
  description of the automerization of cyclobutadiene. \emph{Chem. Phys. Lett.}
  \textbf{2011}, \emph{501}, 166--171\relax
\mciteBstWouldAddEndPuncttrue
\mciteSetBstMidEndSepPunct{\mcitedefaultmidpunct}
{\mcitedefaultendpunct}{\mcitedefaultseppunct}\relax
\EndOfBibitem
\bibitem[Melnichuk and Bartlett(2012)Melnichuk, and Bartlett]{Melnichuk2012}
Melnichuk,~A.; Bartlett,~R.~J. Relaxed active space: Fixing tailored-{CC} with
  high order coupled cluster. I. \emph{J. Chem. Phys.} \textbf{2012},
  \emph{137}, 214103\relax
\mciteBstWouldAddEndPuncttrue
\mciteSetBstMidEndSepPunct{\mcitedefaultmidpunct}
{\mcitedefaultendpunct}{\mcitedefaultseppunct}\relax
\EndOfBibitem
\bibitem[Melnichuk and Bartlett(2014)Melnichuk, and Bartlett]{Melnichuk2014}
Melnichuk,~A.; Bartlett,~R.~J. Relaxed active space: Fixing tailored-{CC} with
  high order coupled cluster. {II}. \emph{J. Chem. Phys.} \textbf{2014},
  \emph{140}, 064113\relax
\mciteBstWouldAddEndPuncttrue
\mciteSetBstMidEndSepPunct{\mcitedefaultmidpunct}
{\mcitedefaultendpunct}{\mcitedefaultseppunct}\relax
\EndOfBibitem
\bibitem[Piecuch \latin{et~al.}(1993)Piecuch, Oliphant, and
  Adamowicz]{Piecuch1993}
Piecuch,~P.; Oliphant,~N.; Adamowicz,~L. A state-selective multireference
  coupled-cluster theory employing the single-reference formalism. \emph{J.
  Chem. Phys.} \textbf{1993}, \emph{99}, 1875--1900\relax
\mciteBstWouldAddEndPuncttrue
\mciteSetBstMidEndSepPunct{\mcitedefaultmidpunct}
{\mcitedefaultendpunct}{\mcitedefaultseppunct}\relax
\EndOfBibitem
\bibitem[Piecuch and Adamowicz(1994)Piecuch, and Adamowicz]{Piecuch1994}
Piecuch,~P.; Adamowicz,~L. State-selective multireference coupled-cluster
  theory employing the single-reference formalism: Implementation and
  application to the H8 model system. \emph{J. Chem. Phys.} \textbf{1994},
  \emph{100}, 5792--5809\relax
\mciteBstWouldAddEndPuncttrue
\mciteSetBstMidEndSepPunct{\mcitedefaultmidpunct}
{\mcitedefaultendpunct}{\mcitedefaultseppunct}\relax
\EndOfBibitem
\bibitem[Adamowicz \latin{et~al.}(1998)Adamowicz, Piecuch, and
  Ghose]{Adamowicz1998}
Adamowicz,~L.; Piecuch,~P.; Ghose,~K.~B. The state-selective coupled cluster
  method for quasi-degenerate electronic states. \emph{Mol. Phys.}
  \textbf{1998}, \emph{94}, 225--234\relax
\mciteBstWouldAddEndPuncttrue
\mciteSetBstMidEndSepPunct{\mcitedefaultmidpunct}
{\mcitedefaultendpunct}{\mcitedefaultseppunct}\relax
\EndOfBibitem
\bibitem[Piecuch(2010)]{Piecuch2010}
Piecuch,~P. Active-space coupled-cluster methods. \emph{Mol. Phys.}
  \textbf{2010}, \emph{108}, 2987--3015\relax
\mciteBstWouldAddEndPuncttrue
\mciteSetBstMidEndSepPunct{\mcitedefaultmidpunct}
{\mcitedefaultendpunct}{\mcitedefaultseppunct}\relax
\EndOfBibitem
\bibitem[Piecuch \latin{et~al.}(2002)Piecuch, Kowalski, Pimienta, and
  Mcguire]{Piecuch2002_1}
Piecuch,~P.; Kowalski,~K.; Pimienta,~I. S.~O.; Mcguire,~M.~J. Recent advances
  in electronic structure theory: Method of moments of coupled-cluster
  equations and renormalized coupled-cluster approaches. \emph{Int. Rev. Phys.
  Chem.} \textbf{2002}, \emph{21}, 527--655\relax
\mciteBstWouldAddEndPuncttrue
\mciteSetBstMidEndSepPunct{\mcitedefaultmidpunct}
{\mcitedefaultendpunct}{\mcitedefaultseppunct}\relax
\EndOfBibitem
\bibitem[Piecuch \latin{et~al.}(2002)Piecuch, Kowalski, and
  Pimienta]{Piecuch2002_2}
Piecuch,~P.; Kowalski,~K.; Pimienta,~I. Method of Moments of Coupled-Cluster
  Equations: Externally Corrected Approaches Employing Configuration
  Interaction Wave Functions. \emph{Int. J. Mol. Sci.} \textbf{2002}, \emph{3},
  475--497\relax
\mciteBstWouldAddEndPuncttrue
\mciteSetBstMidEndSepPunct{\mcitedefaultmidpunct}
{\mcitedefaultendpunct}{\mcitedefaultseppunct}\relax
\EndOfBibitem
\bibitem[Kowalski and Piecuch(2002)Kowalski, and Piecuch]{Kowalski2002}
Kowalski,~K.; Piecuch,~P. Extension of the method of moments of coupled-cluster
  equations to excited states: The triples and quadruples corrections to the
  equation-of-motion coupled-cluster singles and doubles energies. \emph{J.
  Chem. Phys.} \textbf{2002}, \emph{116}, 7411--7423\relax
\mciteBstWouldAddEndPuncttrue
\mciteSetBstMidEndSepPunct{\mcitedefaultmidpunct}
{\mcitedefaultendpunct}{\mcitedefaultseppunct}\relax
\EndOfBibitem
\bibitem[{\l}och \latin{et~al.}(2006){\l}och, Lodriguito, Piecuch{\textdagger},
  and Gour]{Wloch2006}
{\l}och,~M.~W.; Lodriguito,~M.~D.; Piecuch{\textdagger},~P.; Gour,~J.~R. Two
  new classes of non-iterative coupled-cluster methods derived from the method
  of moments of coupled-cluster equations. \emph{Mol. Phys.} \textbf{2006},
  \emph{104}, 2149--2172\relax
\mciteBstWouldAddEndPuncttrue
\mciteSetBstMidEndSepPunct{\mcitedefaultmidpunct}
{\mcitedefaultendpunct}{\mcitedefaultseppunct}\relax
\EndOfBibitem
\bibitem[Lodriguito \latin{et~al.}(2006)Lodriguito, Kowalski, W{\l}och, and
  Piecuch]{Lodriguito2006}
Lodriguito,~M.~D.; Kowalski,~K.; W{\l}och,~M.; Piecuch,~P. Non-iterative
  coupled-cluster methods employing multi-reference perturbation theory wave
  functions. \emph{J. Mol. Struct.: THEOCHEM} \textbf{2006}, \emph{771},
  89--104\relax
\mciteBstWouldAddEndPuncttrue
\mciteSetBstMidEndSepPunct{\mcitedefaultmidpunct}
{\mcitedefaultendpunct}{\mcitedefaultseppunct}\relax
\EndOfBibitem
\bibitem[Piecuch \latin{et~al.}(2004)Piecuch, Kowalski, Pimienta, Fan,
  Lodriguito, McGuire, Kucharski, Ku{\'{s}}, and Musia{\l}]{Piecuch2004}
Piecuch,~P.; Kowalski,~K.; Pimienta,~I. S.~O.; Fan,~P.-D.; Lodriguito,~M.;
  McGuire,~M.~J.; Kucharski,~S.~A.; Ku{\'{s}},~T.; Musia{\l},~M. Method of
  moments of coupled-cluster equations: a new formalism for designing accurate
  electronic structure methods for ground and excited states. \emph{Theor.
  Chem. Acc.} \textbf{2004}, \emph{112}, 349--393\relax
\mciteBstWouldAddEndPuncttrue
\mciteSetBstMidEndSepPunct{\mcitedefaultmidpunct}
{\mcitedefaultendpunct}{\mcitedefaultseppunct}\relax
\EndOfBibitem
\bibitem[Kowalski and Piecuch(2001)Kowalski, and Piecuch]{Kowalski2001}
Kowalski,~K.; Piecuch,~P. New type of noniterative energy corrections for
  excited electronic states: Extension of the method of moments of
  coupled-cluster equations to the equation-of-motion coupled-cluster
  formalism. \emph{J. Chem. Phys.} \textbf{2001}, \emph{115}, 2966--2978\relax
\mciteBstWouldAddEndPuncttrue
\mciteSetBstMidEndSepPunct{\mcitedefaultmidpunct}
{\mcitedefaultendpunct}{\mcitedefaultseppunct}\relax
\EndOfBibitem
\bibitem[Kowalski and Piecuch(2000)Kowalski, and Piecuch]{Kowalski2000}
Kowalski,~K.; Piecuch,~P. The method of moments of coupled-cluster equations
  and the renormalized {CCSD}[T], {CCSD}(T), {CCSD}({TQ}), and {CCSDT}(Q)
  approaches. \emph{J. Chem. Phys.} \textbf{2000}, \emph{113}, 18--35\relax
\mciteBstWouldAddEndPuncttrue
\mciteSetBstMidEndSepPunct{\mcitedefaultmidpunct}
{\mcitedefaultendpunct}{\mcitedefaultseppunct}\relax
\EndOfBibitem
\bibitem[Veis \latin{et~al.}(2016)Veis, Antal{\'{\i}}k, Brabec, Neese, \"{O}rs
  Legeza, and Pittner]{Veis2016}
Veis,~L.; Antal{\'{\i}}k,~A.; Brabec,~J.; Neese,~F.; \"{O}rs Legeza,;
  Pittner,~J. Coupled Cluster Method with Single and Double Excitations
  Tailored by Matrix Product State Wave Functions. \emph{J. Phys. Chem. Lett.}
  \textbf{2016}, \emph{7}, 4072--4078\relax
\mciteBstWouldAddEndPuncttrue
\mciteSetBstMidEndSepPunct{\mcitedefaultmidpunct}
{\mcitedefaultendpunct}{\mcitedefaultseppunct}\relax
\EndOfBibitem
\bibitem[Veis \latin{et~al.}(2017)Veis, Antal{\'{\i}}k, Brabec, Neese, \"{O}rs
  Legeza, and Pittner]{Veis2016_corr}
Veis,~L.; Antal{\'{\i}}k,~A.; Brabec,~J.; Neese,~F.; \"{O}rs Legeza,;
  Pittner,~J. Correction to Coupled Cluster Method with Single and Double
  Excitations Tailored by Matrix Product State Wave Functions. \emph{J. Phys.
  Chem. Lett.} \textbf{2017}, \emph{8}, 291--291\relax
\mciteBstWouldAddEndPuncttrue
\mciteSetBstMidEndSepPunct{\mcitedefaultmidpunct}
{\mcitedefaultendpunct}{\mcitedefaultseppunct}\relax
\EndOfBibitem
\bibitem[Li and Paldus(1997)Li, and Paldus]{paldus-externalcorr}
Li,~X.; Paldus,~J. Reduced multireference CCSD method: An effective approach to
  quasidegenerate states. \emph{J. Chem. Phys} \textbf{1997}, \emph{107},
  6257\relax
\mciteBstWouldAddEndPuncttrue
\mciteSetBstMidEndSepPunct{\mcitedefaultmidpunct}
{\mcitedefaultendpunct}{\mcitedefaultseppunct}\relax
\EndOfBibitem
\bibitem[Deustua \latin{et~al.}(2018)Deustua, Magoulas, Shen, and
  Piecuch]{piecuch-ccmc}
Deustua,~J.~E.; Magoulas,~I.; Shen,~J.; Piecuch,~P. Communication: Approaching
  exact quantum chemistry by cluster analysis of full configuration interaction
  quantum Monte Carlo wave functions. \emph{J. Chem. Phys.} \textbf{2018},
  \emph{149}, 151101\relax
\mciteBstWouldAddEndPuncttrue
\mciteSetBstMidEndSepPunct{\mcitedefaultmidpunct}
{\mcitedefaultendpunct}{\mcitedefaultseppunct}\relax
\EndOfBibitem
\bibitem[White and Noack(1992)White, and Noack]{White1992a}
White,~S.~R.; Noack,~R.~M. Real-space quantum renormalization groups.
  \emph{Phys. Rev. Lett.} \textbf{1992}, \emph{68}, 3487--3490\relax
\mciteBstWouldAddEndPuncttrue
\mciteSetBstMidEndSepPunct{\mcitedefaultmidpunct}
{\mcitedefaultendpunct}{\mcitedefaultseppunct}\relax
\EndOfBibitem
\bibitem[White(1992)]{White1992b}
White,~S.~R. Density matrix formulation for quantum renormalization groups.
  \emph{Phys. Rev. Lett.} \textbf{1992}, \emph{69}, 2863--2866\relax
\mciteBstWouldAddEndPuncttrue
\mciteSetBstMidEndSepPunct{\mcitedefaultmidpunct}
{\mcitedefaultendpunct}{\mcitedefaultseppunct}\relax
\EndOfBibitem
\bibitem[White(1993)]{White1993}
White,~S.~R. Density-matrix algorithms for quantum renormalization groups.
  \emph{Phys. Rev. B} \textbf{1993}, \emph{48}, 10345--10356\relax
\mciteBstWouldAddEndPuncttrue
\mciteSetBstMidEndSepPunct{\mcitedefaultmidpunct}
{\mcitedefaultendpunct}{\mcitedefaultseppunct}\relax
\EndOfBibitem
\bibitem[White and Martin(1999)White, and Martin]{White1999}
White,~S.~R.; Martin,~R.~L. Ab initio quantum chemistry using the density
  matrix renormalization group. \emph{J. Chem. Phys.} \textbf{1999},
  \emph{110}, 4127--4130\relax
\mciteBstWouldAddEndPuncttrue
\mciteSetBstMidEndSepPunct{\mcitedefaultmidpunct}
{\mcitedefaultendpunct}{\mcitedefaultseppunct}\relax
\EndOfBibitem
\bibitem[Chan and Head-Gordon(2002)Chan, and Head-Gordon]{Chan2002a}
Chan,~G. K.-L.; Head-Gordon,~M. Highly correlated calculations with a
  polynomial cost algorithm: A study of the density matrix renormalization
  group. \emph{J. Chem. Phys.} \textbf{2002}, \emph{116}, 4462--4476\relax
\mciteBstWouldAddEndPuncttrue
\mciteSetBstMidEndSepPunct{\mcitedefaultmidpunct}
{\mcitedefaultendpunct}{\mcitedefaultseppunct}\relax
\EndOfBibitem
\bibitem[Legeza \latin{et~al.}(2003)Legeza, R\"{o}der, and Hess]{Legeza2003a}
Legeza,~O.; R\"{o}der,~J.; Hess,~B.~A. Controlling the accuracy of the
  density-matrix renormalization-group method: The dynamical block state
  selection approach. \emph{Phys. Rev. B} \textbf{2003}, \emph{67}\relax
\mciteBstWouldAddEndPuncttrue
\mciteSetBstMidEndSepPunct{\mcitedefaultmidpunct}
{\mcitedefaultendpunct}{\mcitedefaultseppunct}\relax
\EndOfBibitem
\bibitem[Legeza \latin{et~al.}()Legeza, Noack, S{\'{o}}lyom, and
  Tincani]{Legeza2008}
Legeza,~O.; Noack,~R.; S{\'{o}}lyom,~J.; Tincani,~L. \emph{Computational
  Many-Particle Physics}; Springer Berlin Heidelberg, pp 653--664\relax
\mciteBstWouldAddEndPuncttrue
\mciteSetBstMidEndSepPunct{\mcitedefaultmidpunct}
{\mcitedefaultendpunct}{\mcitedefaultseppunct}\relax
\EndOfBibitem
\bibitem[Marti and Reiher(2010)Marti, and Reiher]{Marti2010c}
Marti,~K.~H.; Reiher,~M. The Density Matrix Renormalization Group Algorithm in
  Quantum Chemistry. \emph{Z. Phys. Chem.} \textbf{2010}, \emph{224},
  583--599\relax
\mciteBstWouldAddEndPuncttrue
\mciteSetBstMidEndSepPunct{\mcitedefaultmidpunct}
{\mcitedefaultendpunct}{\mcitedefaultseppunct}\relax
\EndOfBibitem
\bibitem[Chan and Sharma(2011)Chan, and Sharma]{Chan2011}
Chan,~G. K.-L.; Sharma,~S. The Density Matrix Renormalization Group in Quantum
  Chemistry. \emph{Annu. Rev. Phys. Chem.} \textbf{2011}, \emph{62},
  465--481\relax
\mciteBstWouldAddEndPuncttrue
\mciteSetBstMidEndSepPunct{\mcitedefaultmidpunct}
{\mcitedefaultendpunct}{\mcitedefaultseppunct}\relax
\EndOfBibitem
\bibitem[Wouters and Neck(2014)Wouters, and Neck]{Wouters2014_rev}
Wouters,~S.; Neck,~D.~V. The density matrix renormalization group for ab initio
  quantum chemistry. \emph{Eur. Phys. J. D} \textbf{2014}, \emph{68}\relax
\mciteBstWouldAddEndPuncttrue
\mciteSetBstMidEndSepPunct{\mcitedefaultmidpunct}
{\mcitedefaultendpunct}{\mcitedefaultseppunct}\relax
\EndOfBibitem
\bibitem[Szalay \latin{et~al.}(2015)Szalay, Pfeffer, Murg, Barcza, Verstraete,
  Schneider, and \"{O}rs Legeza]{Szalay2015}
Szalay,~S.; Pfeffer,~M.; Murg,~V.; Barcza,~G.; Verstraete,~F.; Schneider,~R.;
  \"{O}rs Legeza, Tensor product methods and entanglement optimization forab
  initioquantum chemistry. \emph{Int. J. Quantum Chem.} \textbf{2015},
  \emph{115}, 1342--1391\relax
\mciteBstWouldAddEndPuncttrue
\mciteSetBstMidEndSepPunct{\mcitedefaultmidpunct}
{\mcitedefaultendpunct}{\mcitedefaultseppunct}\relax
\EndOfBibitem
\bibitem[Yanai \latin{et~al.}(2014)Yanai, Kurashige, Mizukami, Chalupsk{\'{y}},
  Lan, and Saitow]{Yanai2014}
Yanai,~T.; Kurashige,~Y.; Mizukami,~W.; Chalupsk{\'{y}},~J.; Lan,~T.~N.;
  Saitow,~M. Density matrix renormalization group forab {initioCalculations}
  and associated dynamic correlation methods: A review of theory and
  applications. \emph{Int. J. Quantum Chem.} \textbf{2014}, \emph{115},
  283--299\relax
\mciteBstWouldAddEndPuncttrue
\mciteSetBstMidEndSepPunct{\mcitedefaultmidpunct}
{\mcitedefaultendpunct}{\mcitedefaultseppunct}\relax
\EndOfBibitem
\bibitem[Kurashige and Yanai(2011)Kurashige, and Yanai]{Kurashige2011}
Kurashige,~Y.; Yanai,~T. Second-order perturbation theory with a density matrix
  renormalization group self-consistent field reference function: Theory and
  application to the study of chromium dimer. \emph{J. Chem. Phys.}
  \textbf{2011}, \emph{135}, 094104\relax
\mciteBstWouldAddEndPuncttrue
\mciteSetBstMidEndSepPunct{\mcitedefaultmidpunct}
{\mcitedefaultendpunct}{\mcitedefaultseppunct}\relax
\EndOfBibitem
\bibitem[Freitag \latin{et~al.}(2017)Freitag, Knecht, Angeli, and
  Reiher]{Freitag2017}
Freitag,~L.; Knecht,~S.; Angeli,~C.; Reiher,~M. Multireference Perturbation
  Theory with Cholesky Decomposition for the Density Matrix Renormalization
  Group. \emph{J. Chem. Theory Comput.} \textbf{2017}, \emph{13},
  451--459\relax
\mciteBstWouldAddEndPuncttrue
\mciteSetBstMidEndSepPunct{\mcitedefaultmidpunct}
{\mcitedefaultendpunct}{\mcitedefaultseppunct}\relax
\EndOfBibitem
\bibitem[Saitow \latin{et~al.}(2013)Saitow, Kurashige, and Yanai]{Saitow2013}
Saitow,~M.; Kurashige,~Y.; Yanai,~T. Multireference configuration interaction
  theory using cumulant reconstruction with internal contraction of density
  matrix renormalization group wave function. \emph{J. Chem. Phys.}
  \textbf{2013}, \emph{139}, 044118\relax
\mciteBstWouldAddEndPuncttrue
\mciteSetBstMidEndSepPunct{\mcitedefaultmidpunct}
{\mcitedefaultendpunct}{\mcitedefaultseppunct}\relax
\EndOfBibitem
\bibitem[Neuscamman \latin{et~al.}(2010)Neuscamman, Yanai, and
  Chan]{Neuscamman2010}
Neuscamman,~E.; Yanai,~T.; Chan,~G. K.-L. A review of canonical transformation
  theory. \emph{Int. Rev. Phys. Chem.} \textbf{2010}, \emph{29}, 231--271\relax
\mciteBstWouldAddEndPuncttrue
\mciteSetBstMidEndSepPunct{\mcitedefaultmidpunct}
{\mcitedefaultendpunct}{\mcitedefaultseppunct}\relax
\EndOfBibitem
\bibitem[Sharma and Chan(2014)Sharma, and Chan]{Sharma2014}
Sharma,~S.; Chan,~G. K.-L. Communication: A flexible multi-reference
  perturbation theory by minimizing the Hylleraas functional with matrix
  product states. \emph{J. Chem. Phys.} \textbf{2014}, \emph{141}, 111101\relax
\mciteBstWouldAddEndPuncttrue
\mciteSetBstMidEndSepPunct{\mcitedefaultmidpunct}
{\mcitedefaultendpunct}{\mcitedefaultseppunct}\relax
\EndOfBibitem
\bibitem[Sharma \latin{et~al.}(2019)Sharma, Bernales, Knecht, Truhlar, and
  Gagliardi]{Sharma2019}
Sharma,~P.; Bernales,~V.; Knecht,~S.; Truhlar,~D.~G.; Gagliardi,~L. Density
  matrix renormalization group pair-density functional theory ({DMRG}-{PDFT}):
  singlet{\textendash}triplet gaps in polyacenes and polyacetylenes.
  \emph{Chem. Sci.} \textbf{2019}, \emph{10}, 1716--1723\relax
\mciteBstWouldAddEndPuncttrue
\mciteSetBstMidEndSepPunct{\mcitedefaultmidpunct}
{\mcitedefaultendpunct}{\mcitedefaultseppunct}\relax
\EndOfBibitem
\bibitem[Faulstich \latin{et~al.}(2019)Faulstich, Laestadius, \"{O}rs Legeza,
  Schneider, and Kvaal]{Faulstich2019}
Faulstich,~F.~M.; Laestadius,~A.; \"{O}rs Legeza,; Schneider,~R.; Kvaal,~S.
  Analysis of the Tailored Coupled-Cluster Method in Quantum Chemistry.
  \emph{SIAM J. Numer. Anal.} \textbf{2019}, \emph{57}, 2579--2607\relax
\mciteBstWouldAddEndPuncttrue
\mciteSetBstMidEndSepPunct{\mcitedefaultmidpunct}
{\mcitedefaultendpunct}{\mcitedefaultseppunct}\relax
\EndOfBibitem
\bibitem[Faulstich \latin{et~al.}(2019)Faulstich, M{\'{a}}t{\'{e}}, Laestadius,
  Csirik, Veis, Antalik, Brabec, Schneider, Pittner, Kvaal, and \"{O}rs
  Legeza]{Faulstich2019_n2}
Faulstich,~F.~M.; M{\'{a}}t{\'{e}},~M.; Laestadius,~A.; Csirik,~M.~A.;
  Veis,~L.; Antalik,~A.; Brabec,~J.; Schneider,~R.; Pittner,~J.; Kvaal,~S.;
  \"{O}rs Legeza, Numerical and Theoretical Aspects of the {DMRG}-{TCC} Method
  Exemplified by the Nitrogen Dimer. \emph{J. Chem. Theory Comput.}
  \textbf{2019}, \emph{15}, 2206--2220\relax
\mciteBstWouldAddEndPuncttrue
\mciteSetBstMidEndSepPunct{\mcitedefaultmidpunct}
{\mcitedefaultendpunct}{\mcitedefaultseppunct}\relax
\EndOfBibitem
\bibitem[Veis \latin{et~al.}(2018)Veis, Antal{\'{\i}}k, \"{O}rs Legeza, Alavi,
  and Pittner]{Veis2018}
Veis,~L.; Antal{\'{\i}}k,~A.; \"{O}rs Legeza,; Alavi,~A.; Pittner,~J. The
  Intricate Case of Tetramethyleneethane: A Full Configuration Interaction
  Quantum Monte Carlo Benchmark and Multireference Coupled Cluster Studies.
  \emph{J. Chem. Theory Comput.} \textbf{2018}, \emph{14}, 2439--2445\relax
\mciteBstWouldAddEndPuncttrue
\mciteSetBstMidEndSepPunct{\mcitedefaultmidpunct}
{\mcitedefaultendpunct}{\mcitedefaultseppunct}\relax
\EndOfBibitem
\bibitem[Pulay(1983)]{Pulay1983}
Pulay,~P. Localizability of dynamic electron correlation. \emph{Chem. Phys.
  Lett.} \textbf{1983}, \emph{100}, 151--154\relax
\mciteBstWouldAddEndPuncttrue
\mciteSetBstMidEndSepPunct{\mcitedefaultmidpunct}
{\mcitedefaultendpunct}{\mcitedefaultseppunct}\relax
\EndOfBibitem
\bibitem[S{\ae}b{\o} and Pulay(1985)S{\ae}b{\o}, and Pulay]{Saebo1985}
S{\ae}b{\o},~S.; Pulay,~P. Local configuration interaction: An efficient
  approach for larger molecules. \emph{Chem. Phys. Lett.} \textbf{1985},
  \emph{113}, 13--18\relax
\mciteBstWouldAddEndPuncttrue
\mciteSetBstMidEndSepPunct{\mcitedefaultmidpunct}
{\mcitedefaultendpunct}{\mcitedefaultseppunct}\relax
\EndOfBibitem
\bibitem[Foster and Boys(1960)Foster, and Boys]{Foster1960}
Foster,~J.~M.; Boys,~S.~F. Canonical Configurational Interaction Procedure.
  \emph{Rev. Mod. Phys.} \textbf{1960}, \emph{32}, 300--302\relax
\mciteBstWouldAddEndPuncttrue
\mciteSetBstMidEndSepPunct{\mcitedefaultmidpunct}
{\mcitedefaultendpunct}{\mcitedefaultseppunct}\relax
\EndOfBibitem
\bibitem[Pipek and Mezey(1989)Pipek, and Mezey]{Pipek1989}
Pipek,~J.; Mezey,~P.~G. A fast intrinsic localization procedure applicable for
  ab initio and semiempirical linear combination of atomic orbital wave
  functions. \emph{J. Chem. Phys.} \textbf{1989}, \emph{90}, 4916--4926\relax
\mciteBstWouldAddEndPuncttrue
\mciteSetBstMidEndSepPunct{\mcitedefaultmidpunct}
{\mcitedefaultendpunct}{\mcitedefaultseppunct}\relax
\EndOfBibitem
\bibitem[Knizia(2013)]{Knizia2013}
Knizia,~G. Intrinsic Atomic Orbitals: An Unbiased Bridge between Quantum Theory
  and Chemical Concepts. \emph{J. Chem. Theory Comput.} \textbf{2013},
  \emph{9}, 4834--4843\relax
\mciteBstWouldAddEndPuncttrue
\mciteSetBstMidEndSepPunct{\mcitedefaultmidpunct}
{\mcitedefaultendpunct}{\mcitedefaultseppunct}\relax
\EndOfBibitem
\bibitem[S{\ae}b{\o} and Pulay(1987)S{\ae}b{\o}, and Pulay]{Saebo1987}
S{\ae}b{\o},~S.; Pulay,~P. Fourth-order Mo/ller{\textendash}Plessett
  perturbation theory in the local correlation treatment. I. Method. \emph{J.
  Chem. Phys.} \textbf{1987}, \emph{86}, 914--922\relax
\mciteBstWouldAddEndPuncttrue
\mciteSetBstMidEndSepPunct{\mcitedefaultmidpunct}
{\mcitedefaultendpunct}{\mcitedefaultseppunct}\relax
\EndOfBibitem
\bibitem[S{\ae}b{\o} and Pulay(1988)S{\ae}b{\o}, and Pulay]{Saebo1988}
S{\ae}b{\o},~S.; Pulay,~P. The local correlation treatment. {II}.
  Implementation and tests. \emph{J. Chem. Phys.} \textbf{1988}, \emph{88},
  1884--1890\relax
\mciteBstWouldAddEndPuncttrue
\mciteSetBstMidEndSepPunct{\mcitedefaultmidpunct}
{\mcitedefaultendpunct}{\mcitedefaultseppunct}\relax
\EndOfBibitem
\bibitem[Hampel and Werner(1996)Hampel, and Werner]{Hampel1996}
Hampel,~C.; Werner,~H.-J. Local treatment of electron correlation in coupled
  cluster theory. \emph{J. Chem. Phys.} \textbf{1996}, \emph{104},
  6286--6297\relax
\mciteBstWouldAddEndPuncttrue
\mciteSetBstMidEndSepPunct{\mcitedefaultmidpunct}
{\mcitedefaultendpunct}{\mcitedefaultseppunct}\relax
\EndOfBibitem
\bibitem[Sch\"{u}tz and Werner(2001)Sch\"{u}tz, and Werner]{Schutz2001}
Sch\"{u}tz,~M.; Werner,~H.-J. Low-order scaling local electron correlation
  methods. {IV}. Linear scaling local coupled-cluster ({LCCSD}). \emph{J. Chem.
  Phys.} \textbf{2001}, \emph{114}, 661\relax
\mciteBstWouldAddEndPuncttrue
\mciteSetBstMidEndSepPunct{\mcitedefaultmidpunct}
{\mcitedefaultendpunct}{\mcitedefaultseppunct}\relax
\EndOfBibitem
\bibitem[Sch\"{u}tz(2002)]{Schutz2002b}
Sch\"{u}tz,~M. A new, fast, semi-direct implementation of linear scaling local
  coupled cluster theory. \emph{Phys. Chem. Chem. Phys.} \textbf{2002},
  \emph{4}, 3941--3947\relax
\mciteBstWouldAddEndPuncttrue
\mciteSetBstMidEndSepPunct{\mcitedefaultmidpunct}
{\mcitedefaultendpunct}{\mcitedefaultseppunct}\relax
\EndOfBibitem
\bibitem[Werner and Sch\"{u}tz(2011)Werner, and Sch\"{u}tz]{Werner2011}
Werner,~H.-J.; Sch\"{u}tz,~M. An efficient local coupled cluster method for
  accurate thermochemistry of large systems. \emph{J. Chem. Phys.}
  \textbf{2011}, \emph{135}, 144116\relax
\mciteBstWouldAddEndPuncttrue
\mciteSetBstMidEndSepPunct{\mcitedefaultmidpunct}
{\mcitedefaultendpunct}{\mcitedefaultseppunct}\relax
\EndOfBibitem
\bibitem[Sch\"{u}tz(2002)]{Schutz2002a}
Sch\"{u}tz,~M. Low-order scaling local electron correlation methods. V.
  Connected triples beyond (T): Linear scaling local {CCSDT}-1b. \emph{J. Chem.
  Phys.} \textbf{2002}, \emph{116}, 8772--8785\relax
\mciteBstWouldAddEndPuncttrue
\mciteSetBstMidEndSepPunct{\mcitedefaultmidpunct}
{\mcitedefaultendpunct}{\mcitedefaultseppunct}\relax
\EndOfBibitem
\bibitem[Kristensen \latin{et~al.}(2011)Kristensen, Zi{\'{o}}{\l}kowski,
  Jans{\'{\i}}k, Kj{\ae}rgaard, and J{\o}rgensen]{Kristensen2011}
Kristensen,~K.; Zi{\'{o}}{\l}kowski,~M.; Jans{\'{\i}}k,~B.; Kj{\ae}rgaard,~T.;
  J{\o}rgensen,~P. A Locality Analysis of the
  Divide{\textendash}Expand{\textendash}Consolidate Coupled Cluster Amplitude
  Equations. \emph{J. Chem. Theory Comput.} \textbf{2011}, \emph{7},
  1677--1694\relax
\mciteBstWouldAddEndPuncttrue
\mciteSetBstMidEndSepPunct{\mcitedefaultmidpunct}
{\mcitedefaultendpunct}{\mcitedefaultseppunct}\relax
\EndOfBibitem
\bibitem[H{\o}yvik \latin{et~al.}(2012)H{\o}yvik, Kristensen, Jansik, and
  J{\o}rgensen]{Hoyvik2012}
H{\o}yvik,~I.-M.; Kristensen,~K.; Jansik,~B.; J{\o}rgensen,~P. The
  divide-expand-consolidate family of coupled cluster methods: Numerical
  illustrations using second order M{\o}ller-Plesset perturbation theory.
  \emph{J. Chem. Phys.} \textbf{2012}, \emph{136}, 014105\relax
\mciteBstWouldAddEndPuncttrue
\mciteSetBstMidEndSepPunct{\mcitedefaultmidpunct}
{\mcitedefaultendpunct}{\mcitedefaultseppunct}\relax
\EndOfBibitem
\bibitem[Kobayashi and Nakai(2008)Kobayashi, and Nakai]{Kobayashi2008}
Kobayashi,~M.; Nakai,~H. Extension of linear-scaling divide-and-conquer-based
  correlation method to coupled cluster theory with singles and doubles
  excitations. \emph{J. Chem. Phys.} \textbf{2008}, \emph{129}, 044103\relax
\mciteBstWouldAddEndPuncttrue
\mciteSetBstMidEndSepPunct{\mcitedefaultmidpunct}
{\mcitedefaultendpunct}{\mcitedefaultseppunct}\relax
\EndOfBibitem
\bibitem[Stoll(1992)]{Stoll1992}
Stoll,~H. The correlation energy of crystalline silicon. \emph{Chem. Phys.
  Lett.} \textbf{1992}, \emph{191}, 548--552\relax
\mciteBstWouldAddEndPuncttrue
\mciteSetBstMidEndSepPunct{\mcitedefaultmidpunct}
{\mcitedefaultendpunct}{\mcitedefaultseppunct}\relax
\EndOfBibitem
\bibitem[Rolik and K{\'{a}}llay(2011)Rolik, and K{\'{a}}llay]{Rolik2011}
Rolik,~Z.; K{\'{a}}llay,~M. A general-order local coupled-cluster method based
  on the cluster-in-molecule approach. \emph{J. Chem. Phys.} \textbf{2011},
  \emph{135}, 104111\relax
\mciteBstWouldAddEndPuncttrue
\mciteSetBstMidEndSepPunct{\mcitedefaultmidpunct}
{\mcitedefaultendpunct}{\mcitedefaultseppunct}\relax
\EndOfBibitem
\bibitem[Fedorov and Kitaura(2005)Fedorov, and Kitaura]{Fedorov2005}
Fedorov,~D.~G.; Kitaura,~K. Coupled-cluster theory based upon the fragment
  molecular-orbital method. \emph{J. Chem. Phys.} \textbf{2005}, \emph{123},
  134103\relax
\mciteBstWouldAddEndPuncttrue
\mciteSetBstMidEndSepPunct{\mcitedefaultmidpunct}
{\mcitedefaultendpunct}{\mcitedefaultseppunct}\relax
\EndOfBibitem
\bibitem[Li \latin{et~al.}(2001)Li, Ma, and Jiang]{Li2001}
Li,~S.; Ma,~J.; Jiang,~Y. Linear scaling local correlation approach for solving
  the coupled cluster equations of large systems. \emph{J. Comput. Chem.}
  \textbf{2001}, \emph{23}, 237--244\relax
\mciteBstWouldAddEndPuncttrue
\mciteSetBstMidEndSepPunct{\mcitedefaultmidpunct}
{\mcitedefaultendpunct}{\mcitedefaultseppunct}\relax
\EndOfBibitem
\bibitem[Edmiston and Krauss(1965)Edmiston, and Krauss]{Edmiston1965}
Edmiston,~C.; Krauss,~M. Configuration-Interaction Calculation of H3 and H2.
  \emph{J. Chem. Phys.} \textbf{1965}, \emph{42}, 1119--1120\relax
\mciteBstWouldAddEndPuncttrue
\mciteSetBstMidEndSepPunct{\mcitedefaultmidpunct}
{\mcitedefaultendpunct}{\mcitedefaultseppunct}\relax
\EndOfBibitem
\bibitem[Meyer(1971)]{Meyer1971}
Meyer,~W. Ionization energies of water from {PNO}-{CI} calculations. \emph{Int.
  J. Quantum Chem.} \textbf{1971}, \emph{5}, 341--348\relax
\mciteBstWouldAddEndPuncttrue
\mciteSetBstMidEndSepPunct{\mcitedefaultmidpunct}
{\mcitedefaultendpunct}{\mcitedefaultseppunct}\relax
\EndOfBibitem
\bibitem[Meyer(1974)]{Meyer1974}
Meyer,~W. {PNO}-{CI} and {CEPA} studies of electron correlation effects.
  \emph{Theor. Chim. Acta} \textbf{1974}, \emph{35}, 277--292\relax
\mciteBstWouldAddEndPuncttrue
\mciteSetBstMidEndSepPunct{\mcitedefaultmidpunct}
{\mcitedefaultendpunct}{\mcitedefaultseppunct}\relax
\EndOfBibitem
\bibitem[Werner and Meyer(1976)Werner, and Meyer]{Werner1976}
Werner,~H.-J.; Meyer,~W. {PNO}-{CI} and {PNO}-{CEPA} studies of electron
  correlation effects. \emph{Mol. Phys.} \textbf{1976}, \emph{31},
  855--872\relax
\mciteBstWouldAddEndPuncttrue
\mciteSetBstMidEndSepPunct{\mcitedefaultmidpunct}
{\mcitedefaultendpunct}{\mcitedefaultseppunct}\relax
\EndOfBibitem
\bibitem[Botschwina and Meyer(1977)Botschwina, and Meyer]{Botschwina1977}
Botschwina,~P.; Meyer,~W. {PNO}-{CEPA} calculation of collinear potential
  energy barriers for thermoneutral exchange reactions. \emph{Chem. Phys.}
  \textbf{1977}, \emph{20}, 43--52\relax
\mciteBstWouldAddEndPuncttrue
\mciteSetBstMidEndSepPunct{\mcitedefaultmidpunct}
{\mcitedefaultendpunct}{\mcitedefaultseppunct}\relax
\EndOfBibitem
\bibitem[Rosmus and Meyer(1978)Rosmus, and Meyer]{Rosmus1978}
Rosmus,~P.; Meyer,~W. {PNO}-{CI} and {CEPA} studies of electron correlation
  effects. {VI}. Electron affinities of the first-row and second-row diatomic
  hydrides and the spectroscopic constants of their negative ions. \emph{J.
  Chem. Phys.} \textbf{1978}, \emph{69}, 2745\relax
\mciteBstWouldAddEndPuncttrue
\mciteSetBstMidEndSepPunct{\mcitedefaultmidpunct}
{\mcitedefaultendpunct}{\mcitedefaultseppunct}\relax
\EndOfBibitem
\bibitem[Ahlrichs and Kutzelnigg(1968)Ahlrichs, and Kutzelnigg]{Ahlrichs1968}
Ahlrichs,~R.; Kutzelnigg,~W. Ab initio calculations on small hydrides including
  electron correlation. \emph{Theor. Chim. Acta} \textbf{1968}, \emph{10},
  377--387\relax
\mciteBstWouldAddEndPuncttrue
\mciteSetBstMidEndSepPunct{\mcitedefaultmidpunct}
{\mcitedefaultendpunct}{\mcitedefaultseppunct}\relax
\EndOfBibitem
\bibitem[Ahlrichs \latin{et~al.}(1975)Ahlrichs, Driessler, Lischka, Staemmler,
  and Kutzelnigg]{Ahlrichs1975}
Ahlrichs,~R.; Driessler,~F.; Lischka,~H.; Staemmler,~V.; Kutzelnigg,~W. PNO-CI
  (pair natural orbital configuration interaction) and CEPA-PNO (coupled
  electron pair approximation with pair natural orbitals) calculations of
  molecular systems. {II}. The molecules BeH$_2$, BH, BH$_3$, CH$_4$, CH$_3^-$,
  NH$_3$ (planar and pyramidal), H$_2$O, OH$_3^+$, {HF} and the Ne atom.
  \emph{J. Chem. Phys.} \textbf{1975}, \emph{62}, 1235--1247\relax
\mciteBstWouldAddEndPuncttrue
\mciteSetBstMidEndSepPunct{\mcitedefaultmidpunct}
{\mcitedefaultendpunct}{\mcitedefaultseppunct}\relax
\EndOfBibitem
\bibitem[Neese \latin{et~al.}(2009)Neese, Wennmohs, and Hansen]{Neese2009_cepa}
Neese,~F.; Wennmohs,~F.; Hansen,~A. Efficient and accurate local approximations
  to coupled-electron pair approaches: An attempt to revive the pair natural
  orbital method. \emph{J. Chem. Phys.} \textbf{2009}, \emph{130}, 114108\relax
\mciteBstWouldAddEndPuncttrue
\mciteSetBstMidEndSepPunct{\mcitedefaultmidpunct}
{\mcitedefaultendpunct}{\mcitedefaultseppunct}\relax
\EndOfBibitem
\bibitem[Neese \latin{et~al.}(2009)Neese, Hansen, and Liakos]{Neese2009_ccsd}
Neese,~F.; Hansen,~A.; Liakos,~D.~G. Efficient and accurate approximations to
  the local coupled cluster singles doubles method using a truncated pair
  natural orbital basis. \emph{J. Chem. Phys.} \textbf{2009}, \emph{131},
  064103\relax
\mciteBstWouldAddEndPuncttrue
\mciteSetBstMidEndSepPunct{\mcitedefaultmidpunct}
{\mcitedefaultendpunct}{\mcitedefaultseppunct}\relax
\EndOfBibitem
\bibitem[Hansen \latin{et~al.}(2011)Hansen, Liakos, and Neese]{Hansen2011}
Hansen,~A.; Liakos,~D.~G.; Neese,~F. Efficient and accurate local single
  reference correlation methods for high-spin open-shell molecules using pair
  natural orbitals. \emph{J. Chem. Phys.} \textbf{2011}, \emph{135},
  214102\relax
\mciteBstWouldAddEndPuncttrue
\mciteSetBstMidEndSepPunct{\mcitedefaultmidpunct}
{\mcitedefaultendpunct}{\mcitedefaultseppunct}\relax
\EndOfBibitem
\bibitem[Huntington \latin{et~al.}(2012)Huntington, Hansen, Neese, and
  Nooijen]{Huntington2012}
Huntington,~L. M.~J.; Hansen,~A.; Neese,~F.; Nooijen,~M. Accurate
  thermochemistry from a parameterized coupled-cluster singles and doubles
  model and a local pair natural orbital based implementation for applications
  to larger systems. \emph{J. Chem. Phys.} \textbf{2012}, \emph{136},
  064101\relax
\mciteBstWouldAddEndPuncttrue
\mciteSetBstMidEndSepPunct{\mcitedefaultmidpunct}
{\mcitedefaultendpunct}{\mcitedefaultseppunct}\relax
\EndOfBibitem
\bibitem[Riplinger and Neese(2013)Riplinger, and Neese]{Riplinger2013_dlpno}
Riplinger,~C.; Neese,~F. An efficient and near linear scaling pair natural
  orbital based local coupled cluster method. \emph{J. Chem. Phys.}
  \textbf{2013}, \emph{138}, 034106\relax
\mciteBstWouldAddEndPuncttrue
\mciteSetBstMidEndSepPunct{\mcitedefaultmidpunct}
{\mcitedefaultendpunct}{\mcitedefaultseppunct}\relax
\EndOfBibitem
\bibitem[Pinski \latin{et~al.}(2015)Pinski, Riplinger, Valeev, and
  Neese]{Pinski2015}
Pinski,~P.; Riplinger,~C.; Valeev,~E.~F.; Neese,~F. Sparse maps{\textemdash}A
  systematic infrastructure for reduced-scaling electronic structure methods.
  I. An efficient and simple linear scaling local {MP}2 method that uses an
  intermediate basis of pair natural orbitals. \emph{J. Chem. Phys.}
  \textbf{2015}, \emph{143}, 034108\relax
\mciteBstWouldAddEndPuncttrue
\mciteSetBstMidEndSepPunct{\mcitedefaultmidpunct}
{\mcitedefaultendpunct}{\mcitedefaultseppunct}\relax
\EndOfBibitem
\bibitem[Riplinger \latin{et~al.}(2016)Riplinger, Pinski, Becker, Valeev, and
  Neese]{Riplinger2016}
Riplinger,~C.; Pinski,~P.; Becker,~U.; Valeev,~E.~F.; Neese,~F. Sparse
  maps{\textemdash}A systematic infrastructure for reduced-scaling electronic
  structure methods. {II}. Linear scaling domain based pair natural orbital
  coupled cluster theory. \emph{J. Chem. Phys.} \textbf{2016}, \emph{144},
  024109\relax
\mciteBstWouldAddEndPuncttrue
\mciteSetBstMidEndSepPunct{\mcitedefaultmidpunct}
{\mcitedefaultendpunct}{\mcitedefaultseppunct}\relax
\EndOfBibitem
\bibitem[Saitow \latin{et~al.}(2017)Saitow, Becker, Riplinger, Valeev, and
  Neese]{Saitow2017}
Saitow,~M.; Becker,~U.; Riplinger,~C.; Valeev,~E.~F.; Neese,~F. A new
  near-linear scaling, efficient and accurate, open-shell domain-based local
  pair natural orbital coupled cluster singles and doubles theory. \emph{J.
  Chem. Phys.} \textbf{2017}, \emph{146}, 164105\relax
\mciteBstWouldAddEndPuncttrue
\mciteSetBstMidEndSepPunct{\mcitedefaultmidpunct}
{\mcitedefaultendpunct}{\mcitedefaultseppunct}\relax
\EndOfBibitem
\bibitem[Werner \latin{et~al.}(2015)Werner, Knizia, Krause, Schwilk, and
  Dornbach]{Werner2015}
Werner,~H.-J.; Knizia,~G.; Krause,~C.; Schwilk,~M.; Dornbach,~M. Scalable
  Electron Correlation Methods I.: {PNO}-{LMP}2 with Linear Scaling in the
  Molecular Size and Near-Inverse-Linear Scaling in the Number of Processors.
  \emph{J. Chem. Theory Comput.} \textbf{2015}, \emph{11}, 484--507\relax
\mciteBstWouldAddEndPuncttrue
\mciteSetBstMidEndSepPunct{\mcitedefaultmidpunct}
{\mcitedefaultendpunct}{\mcitedefaultseppunct}\relax
\EndOfBibitem
\bibitem[Menezes \latin{et~al.}(2016)Menezes, Kats, and Werner]{Menezes2016}
Menezes,~F.; Kats,~D.; Werner,~H.-J. Local complete active space second-order
  perturbation theory using pair natural orbitals ({PNO}-{CASPT}2). \emph{J.
  Chem. Phys.} \textbf{2016}, \emph{145}, 124115\relax
\mciteBstWouldAddEndPuncttrue
\mciteSetBstMidEndSepPunct{\mcitedefaultmidpunct}
{\mcitedefaultendpunct}{\mcitedefaultseppunct}\relax
\EndOfBibitem
\bibitem[Schwilk \latin{et~al.}(2017)Schwilk, Ma, K\"{o}ppl, and
  Werner]{Schwilk2017}
Schwilk,~M.; Ma,~Q.; K\"{o}ppl,~C.; Werner,~H.-J. Scalable Electron Correlation
  Methods. 3. Efficient and Accurate Parallel Local Coupled Cluster with Pair
  Natural Orbitals ({PNO}-{LCCSD}). \emph{J. Chem. Theory Comput.}
  \textbf{2017}, \emph{13}, 3650--3675\relax
\mciteBstWouldAddEndPuncttrue
\mciteSetBstMidEndSepPunct{\mcitedefaultmidpunct}
{\mcitedefaultendpunct}{\mcitedefaultseppunct}\relax
\EndOfBibitem
\bibitem[Tew \latin{et~al.}(2011)Tew, Helmich, and H\"{a}ttig]{Tew2011}
Tew,~D.~P.; Helmich,~B.; H\"{a}ttig,~C. Local explicitly correlated
  second-order M{\o}ller{\textendash}Plesset perturbation theory with pair
  natural orbitals. \emph{J. Chem. Phys.} \textbf{2011}, \emph{135},
  074107\relax
\mciteBstWouldAddEndPuncttrue
\mciteSetBstMidEndSepPunct{\mcitedefaultmidpunct}
{\mcitedefaultendpunct}{\mcitedefaultseppunct}\relax
\EndOfBibitem
\bibitem[Helmich and H\"{a}ttig(2013)Helmich, and H\"{a}ttig]{Helmich2013}
Helmich,~B.; H\"{a}ttig,~C. A pair natural orbital implementation of the
  coupled cluster model {CC}2 for excitation energies. \emph{J. Chem. Phys.}
  \textbf{2013}, \emph{139}, 084114\relax
\mciteBstWouldAddEndPuncttrue
\mciteSetBstMidEndSepPunct{\mcitedefaultmidpunct}
{\mcitedefaultendpunct}{\mcitedefaultseppunct}\relax
\EndOfBibitem
\bibitem[Schmitz \latin{et~al.}(2013)Schmitz, Helmich, and
  H\"{a}ttig]{Schmitz2013}
Schmitz,~G.; Helmich,~B.; H\"{a}ttig,~C. A scaling {PNO}{\textendash}{MP}2
  method using a hybrid {OSV}{\textendash}{PNO} approach with an iterative
  direct generation of {OSVs}{\textdagger}. \emph{Mol. Phys.} \textbf{2013},
  \emph{111}, 2463--2476\relax
\mciteBstWouldAddEndPuncttrue
\mciteSetBstMidEndSepPunct{\mcitedefaultmidpunct}
{\mcitedefaultendpunct}{\mcitedefaultseppunct}\relax
\EndOfBibitem
\bibitem[Guo \latin{et~al.}(2016)Guo, Sivalingam, Valeev, and Neese]{Guo2016}
Guo,~Y.; Sivalingam,~K.; Valeev,~E.~F.; Neese,~F. {SparseMaps}{\textemdash}A
  systematic infrastructure for reduced-scaling electronic structure methods.
  {III}. Linear-scaling multireference domain-based pair natural orbital
  N-electron valence perturbation theory. \emph{J. Chem. Phys.} \textbf{2016},
  \emph{144}, 094111\relax
\mciteBstWouldAddEndPuncttrue
\mciteSetBstMidEndSepPunct{\mcitedefaultmidpunct}
{\mcitedefaultendpunct}{\mcitedefaultseppunct}\relax
\EndOfBibitem
\bibitem[Antony \latin{et~al.}(2011)Antony, Grimme, Liakos, and
  Neese]{Antony2011}
Antony,~J.; Grimme,~S.; Liakos,~D.~G.; Neese,~F. Protein{\textendash}Ligand
  Interaction Energies with Dispersion Corrected Density Functional Theory and
  High-Level Wave Function Based Methods. \emph{J. Phys. Chem. A}
  \textbf{2011}, \emph{115}, 11210--11220\relax
\mciteBstWouldAddEndPuncttrue
\mciteSetBstMidEndSepPunct{\mcitedefaultmidpunct}
{\mcitedefaultendpunct}{\mcitedefaultseppunct}\relax
\EndOfBibitem
\bibitem[Anoop \latin{et~al.}(2010)Anoop, Thiel, and Neese]{Anoop2010}
Anoop,~A.; Thiel,~W.; Neese,~F. A Local Pair Natural Orbital Coupled Cluster
  Study of Rh Catalyzed Asymmetric Olefin Hydrogenation. \emph{J. Chem. Theory
  Comput.} \textbf{2010}, \emph{6}, 3137--3144\relax
\mciteBstWouldAddEndPuncttrue
\mciteSetBstMidEndSepPunct{\mcitedefaultmidpunct}
{\mcitedefaultendpunct}{\mcitedefaultseppunct}\relax
\EndOfBibitem
\bibitem[Liakos and Neese(2011)Liakos, and Neese]{Liakos2011}
Liakos,~D.~G.; Neese,~F. Interplay of Correlation and Relativistic Effects in
  Correlated Calculations on Transition-Metal Complexes: The
  (Cu$_2$O$_2$)$^{2+}$ Core Revisited. \emph{J. Chem. Theory Comput.}
  \textbf{2011}, \emph{7}, 1511--1523\relax
\mciteBstWouldAddEndPuncttrue
\mciteSetBstMidEndSepPunct{\mcitedefaultmidpunct}
{\mcitedefaultendpunct}{\mcitedefaultseppunct}\relax
\EndOfBibitem
\bibitem[Zade \latin{et~al.}(2011)Zade, Zamoshchik, Reddy, Fridman-Marueli,
  Sheberla, and Bendikov]{Zade2011}
Zade,~S.~S.; Zamoshchik,~N.; Reddy,~A.~R.; Fridman-Marueli,~G.; Sheberla,~D.;
  Bendikov,~M. Products and Mechanism of Acene Dimerization. A Computational
  Study. \emph{J. Am. Chem. Soc.} \textbf{2011}, \emph{133}, 10803--10816\relax
\mciteBstWouldAddEndPuncttrue
\mciteSetBstMidEndSepPunct{\mcitedefaultmidpunct}
{\mcitedefaultendpunct}{\mcitedefaultseppunct}\relax
\EndOfBibitem
\bibitem[Kubas \latin{et~al.}(2012)Kubas, Br\"{a}se, and Fink]{Kubas2012}
Kubas,~A.; Br\"{a}se,~S.; Fink,~K. Theoretical Approach Towards the
  Understanding of Asymmetric Additions of Dialkylzinc to Enals and Iminals
  Catalysed by [2.2]Paracyclophane-Based N, O-Ligands. \emph{Chem.: Eur. J.}
  \textbf{2012}, \emph{18}, 8377--8385\relax
\mciteBstWouldAddEndPuncttrue
\mciteSetBstMidEndSepPunct{\mcitedefaultmidpunct}
{\mcitedefaultendpunct}{\mcitedefaultseppunct}\relax
\EndOfBibitem
\bibitem[Ashtari and Cann(2011)Ashtari, and Cann]{Ashtari2011}
Ashtari,~M.; Cann,~N. Proline-based chiral stationary phases: A molecular
  dynamics study of the interfacial structure. \emph{J. Chromatogr. A}
  \textbf{2011}, \emph{1218}, 6331--6347\relax
\mciteBstWouldAddEndPuncttrue
\mciteSetBstMidEndSepPunct{\mcitedefaultmidpunct}
{\mcitedefaultendpunct}{\mcitedefaultseppunct}\relax
\EndOfBibitem
\bibitem[Zhang and Dolg(2014)Zhang, and Dolg]{Zhang2014}
Zhang,~J.; Dolg,~M. Dispersion Interaction Stabilizes Sterically Hindered
  Double Fullerenes. \emph{Chem.: Eur. J.} \textbf{2014}, \emph{20},
  13909--13912\relax
\mciteBstWouldAddEndPuncttrue
\mciteSetBstMidEndSepPunct{\mcitedefaultmidpunct}
{\mcitedefaultendpunct}{\mcitedefaultseppunct}\relax
\EndOfBibitem
\bibitem[Minenkov \latin{et~al.}(2015)Minenkov, Chermak, and
  Cavallo]{Minenkov2015}
Minenkov,~Y.; Chermak,~E.; Cavallo,~L. Accuracy of
  {DLPNO}{\textendash}{CCSD}(T) Method for Noncovalent Bond Dissociation
  Enthalpies from Coinage Metal Cation Complexes. \emph{J. Chem. Theory
  Comput.} \textbf{2015}, \emph{11}, 4664--4676\relax
\mciteBstWouldAddEndPuncttrue
\mciteSetBstMidEndSepPunct{\mcitedefaultmidpunct}
{\mcitedefaultendpunct}{\mcitedefaultseppunct}\relax
\EndOfBibitem
\bibitem[Sparta and Neese(2014)Sparta, and Neese]{Sparta2014}
Sparta,~M.; Neese,~F. Chemical applications carried out by local pair natural
  orbital based coupled-cluster methods. \emph{Chem. Soc. Rev.} \textbf{2014},
  \emph{43}, 5032--5041\relax
\mciteBstWouldAddEndPuncttrue
\mciteSetBstMidEndSepPunct{\mcitedefaultmidpunct}
{\mcitedefaultendpunct}{\mcitedefaultseppunct}\relax
\EndOfBibitem
\bibitem[Liakos \latin{et~al.}(2015)Liakos, Sparta, Kesharwani, Martin, and
  Neese]{Liakos2015}
Liakos,~D.~G.; Sparta,~M.; Kesharwani,~M.~K.; Martin,~J. M.~L.; Neese,~F.
  Exploring the Accuracy Limits of Local Pair Natural Orbital Coupled-Cluster
  Theory. \emph{J. Chem. Theory Comput.} \textbf{2015}, \emph{11},
  1525--1539\relax
\mciteBstWouldAddEndPuncttrue
\mciteSetBstMidEndSepPunct{\mcitedefaultmidpunct}
{\mcitedefaultendpunct}{\mcitedefaultseppunct}\relax
\EndOfBibitem
\bibitem[Demel \latin{et~al.}(2015)Demel, Pittner, and Neese]{Demel2015}
Demel,~O.; Pittner,~J.; Neese,~F. A Local Pair Natural Orbital-Based
  Multireference Mukherjee's Coupled Cluster Method. \emph{J. Chem. Theory
  Comput.} \textbf{2015}, \emph{11}, 3104--3114\relax
\mciteBstWouldAddEndPuncttrue
\mciteSetBstMidEndSepPunct{\mcitedefaultmidpunct}
{\mcitedefaultendpunct}{\mcitedefaultseppunct}\relax
\EndOfBibitem
\bibitem[Lang \latin{et~al.}(2017)Lang, {\v{S}}va{\v{n}}a, Demel, Brabec,
  Ked{\v{z}}uch, Noga, Kowalski, and Pittner]{Lang2017}
Lang,~J.; {\v{S}}va{\v{n}}a,~M.; Demel,~O.; Brabec,~J.; Ked{\v{z}}uch,~S.;
  Noga,~J.; Kowalski,~K.; Pittner,~J. A {MRCC} study of the isomerisation of
  cyclopropane. \emph{Mol. Phys.} \textbf{2017}, \emph{115}, 2743--2754\relax
\mciteBstWouldAddEndPuncttrue
\mciteSetBstMidEndSepPunct{\mcitedefaultmidpunct}
{\mcitedefaultendpunct}{\mcitedefaultseppunct}\relax
\EndOfBibitem
\bibitem[Brabec \latin{et~al.}(2018)Brabec, Lang, Saitow, Pittner, Neese, and
  Demel]{Brabec2018}
Brabec,~J.; Lang,~J.; Saitow,~M.; Pittner,~J.; Neese,~F.; Demel,~O.
  Domain-Based Local Pair Natural Orbital Version of Mukherjee's State-Specific
  Coupled Cluster Method. \emph{J. Chem. Theory Comput.} \textbf{2018},
  \emph{14}, 1370--1382\relax
\mciteBstWouldAddEndPuncttrue
\mciteSetBstMidEndSepPunct{\mcitedefaultmidpunct}
{\mcitedefaultendpunct}{\mcitedefaultseppunct}\relax
\EndOfBibitem
\bibitem[Lang \latin{et~al.}(2019)Lang, Brabec, Saitow, Pittner, Neese, and
  Demel]{Lang2019}
Lang,~J.; Brabec,~J.; Saitow,~M.; Pittner,~J.; Neese,~F.; Demel,~O.
  Perturbative triples correction to domain-based local pair natural orbital
  variants of Mukherjee's state specific coupled cluster method. \emph{Phys.
  Chem. Chem. Phys.} \textbf{2019}, \emph{21}, 5022--5038\relax
\mciteBstWouldAddEndPuncttrue
\mciteSetBstMidEndSepPunct{\mcitedefaultmidpunct}
{\mcitedefaultendpunct}{\mcitedefaultseppunct}\relax
\EndOfBibitem
\bibitem[Antal{\'{\i}}k \latin{et~al.}(2019)Antal{\'{\i}}k, Veis, Brabec,
  Demel, \"{O}rs Legeza, and Pittner]{Antalik2019}
Antal{\'{\i}}k,~A.; Veis,~L.; Brabec,~J.; Demel,~O.; \"{O}rs Legeza,;
  Pittner,~J. Toward the efficient local tailored coupled cluster approximation
  and the peculiar case of oxo-Mn(Salen). \emph{J. Chem. Phys.} \textbf{2019},
  \emph{151}, 084112\relax
\mciteBstWouldAddEndPuncttrue
\mciteSetBstMidEndSepPunct{\mcitedefaultmidpunct}
{\mcitedefaultendpunct}{\mcitedefaultseppunct}\relax
\EndOfBibitem
\bibitem[Kowalski(2018)]{Kowalski2018}
Kowalski,~K. Properties of coupled-cluster equations originating in excitation
  sub-algebras. \emph{The Journal of Chemical Physics} \textbf{2018},
  \emph{148}, 094104\relax
\mciteBstWouldAddEndPuncttrue
\mciteSetBstMidEndSepPunct{\mcitedefaultmidpunct}
{\mcitedefaultendpunct}{\mcitedefaultseppunct}\relax
\EndOfBibitem
\bibitem[Bauman \latin{et~al.}(2019)Bauman, Bylaska, Krishnamoorthy, Low,
  Wiebe, Granade, Roetteler, Troyer, and Kowalski]{Bauman2019}
Bauman,~N.~P.; Bylaska,~E.~J.; Krishnamoorthy,~S.; Low,~G.~H.; Wiebe,~N.;
  Granade,~C.~E.; Roetteler,~M.; Troyer,~M.; Kowalski,~K. Downfolding of
  many-body Hamiltonians using active-space models: Extension of the sub-system
  embedding sub-algebras approach to unitary coupled cluster formalisms.
  \emph{The Journal of Chemical Physics} \textbf{2019}, \emph{151},
  014107\relax
\mciteBstWouldAddEndPuncttrue
\mciteSetBstMidEndSepPunct{\mcitedefaultmidpunct}
{\mcitedefaultendpunct}{\mcitedefaultseppunct}\relax
\EndOfBibitem
\bibitem[Moritz and Reiher(2007)Moritz, and Reiher]{Moritz2007}
Moritz,~G.; Reiher,~M. Decomposition of density matrix renormalization group
  states into a Slater determinant basis. \emph{J. Chem. Phys.} \textbf{2007},
  \emph{126}, 244109\relax
\mciteBstWouldAddEndPuncttrue
\mciteSetBstMidEndSepPunct{\mcitedefaultmidpunct}
{\mcitedefaultendpunct}{\mcitedefaultseppunct}\relax
\EndOfBibitem
\bibitem[Boguslawski \latin{et~al.}(2011)Boguslawski, Marti, and
  Reiher]{Boguslawski2011}
Boguslawski,~K.; Marti,~K.~H.; Reiher,~M. Construction of {CASCI}-type wave
  functions for very large active spaces. \emph{J. Chem. Phys.} \textbf{2011},
  \emph{134}, 224101\relax
\mciteBstWouldAddEndPuncttrue
\mciteSetBstMidEndSepPunct{\mcitedefaultmidpunct}
{\mcitedefaultendpunct}{\mcitedefaultseppunct}\relax
\EndOfBibitem
\bibitem[Neese(2011)]{Neese2011}
Neese,~F. The {ORCA} program system. \emph{Wiley Interdiscip. Rev. Comput. Mol.
  Sci.} \textbf{2011}, \emph{2}, 73--78\relax
\mciteBstWouldAddEndPuncttrue
\mciteSetBstMidEndSepPunct{\mcitedefaultmidpunct}
{\mcitedefaultendpunct}{\mcitedefaultseppunct}\relax
\EndOfBibitem
\bibitem[Olivares-Amaya \latin{et~al.}(2015)Olivares-Amaya, Hu, Nakatani,
  Sharma, Yang, and Chan]{OlivaresAmaya2015}
Olivares-Amaya,~R.; Hu,~W.; Nakatani,~N.; Sharma,~S.; Yang,~J.; Chan,~G. K.-L.
  The ab-initio density matrix renormalization group in practice. \emph{J.
  Chem. Phys.} \textbf{2015}, \emph{142}, 034102\relax
\mciteBstWouldAddEndPuncttrue
\mciteSetBstMidEndSepPunct{\mcitedefaultmidpunct}
{\mcitedefaultendpunct}{\mcitedefaultseppunct}\relax
\EndOfBibitem
\bibitem[Adler and Werner(2011)Adler, and Werner]{Adler2011}
Adler,~T.~B.; Werner,~H.-J. An explicitly correlated local coupled cluster
  method for calculations of large molecules close to the basis set limit.
  \emph{J. Chem. Phys.} \textbf{2011}, \emph{135}, 144117\relax
\mciteBstWouldAddEndPuncttrue
\mciteSetBstMidEndSepPunct{\mcitedefaultmidpunct}
{\mcitedefaultendpunct}{\mcitedefaultseppunct}\relax
\EndOfBibitem
\bibitem[Adler and Werner(2009)Adler, and Werner]{Adler2009}
Adler,~T.~B.; Werner,~H.-J. Local explicitly correlated coupled-cluster
  methods: Efficient removal of the basis set incompleteness and domain errors.
  \emph{J. Chem. Phys.} \textbf{2009}, \emph{130}, 241101\relax
\mciteBstWouldAddEndPuncttrue
\mciteSetBstMidEndSepPunct{\mcitedefaultmidpunct}
{\mcitedefaultendpunct}{\mcitedefaultseppunct}\relax
\EndOfBibitem
\bibitem[Riplinger \latin{et~al.}(2013)Riplinger, Sandhoefer, Hansen, and
  Neese]{Riplinger2013_triples}
Riplinger,~C.; Sandhoefer,~B.; Hansen,~A.; Neese,~F. Natural triple excitations
  in local coupled cluster calculations with pair natural orbitals. \emph{J.
  Chem. Phys.} \textbf{2013}, \emph{139}, 134101\relax
\mciteBstWouldAddEndPuncttrue
\mciteSetBstMidEndSepPunct{\mcitedefaultmidpunct}
{\mcitedefaultendpunct}{\mcitedefaultseppunct}\relax
\EndOfBibitem
\bibitem[Guo \latin{et~al.}(2018)Guo, Riplinger, Becker, Liakos, Minenkov,
  Cavallo, and Neese]{Guo2018}
Guo,~Y.; Riplinger,~C.; Becker,~U.; Liakos,~D.~G.; Minenkov,~Y.; Cavallo,~L.;
  Neese,~F. Communication: An improved linear scaling perturbative triples
  correction for the domain based local pair-natural orbital based singles and
  doubles coupled cluster method [{DLPNO}-{CCSD}(T)]. \emph{J. Chem. Phys.}
  \textbf{2018}, \emph{148}, 011101\relax
\mciteBstWouldAddEndPuncttrue
\mciteSetBstMidEndSepPunct{\mcitedefaultmidpunct}
{\mcitedefaultendpunct}{\mcitedefaultseppunct}\relax
\EndOfBibitem
\bibitem[Legeza \latin{et~al.}()Legeza, Veis, and Mosoni]{budapest_qcdmrg}
Legeza,~{\"O}.; Veis,~L.; Mosoni,~T. {QC-DMRG-Budapest, a program for quantum
  chemical DMRG calculations}\relax
\mciteBstWouldAddEndPuncttrue
\mciteSetBstMidEndSepPunct{\mcitedefaultmidpunct}
{\mcitedefaultendpunct}{\mcitedefaultseppunct}\relax
\EndOfBibitem
\bibitem[Barcza \latin{et~al.}(2011)Barcza, Legeza, Marti, and
  Reiher]{Barcza2011}
Barcza,~G.; Legeza,~O.; Marti,~K.~H.; Reiher,~M. Quantum-information analysis
  of electronic states of different molecular structures. \emph{Phys. Rev. A}
  \textbf{2011}, \emph{83}\relax
\mciteBstWouldAddEndPuncttrue
\mciteSetBstMidEndSepPunct{\mcitedefaultmidpunct}
{\mcitedefaultendpunct}{\mcitedefaultseppunct}\relax
\EndOfBibitem
\bibitem[Fertitta \latin{et~al.}(2014)Fertitta, Paulus, Barcza, and
  Legeza]{Fertitta2014}
Fertitta,~E.; Paulus,~B.; Barcza,~G.; Legeza,~O. Investigation of
  metal{\textendash}insulator-like transition through theab initiodensity
  matrix renormalization group approach. \emph{Phys. Rev. B} \textbf{2014},
  \emph{90}\relax
\mciteBstWouldAddEndPuncttrue
\mciteSetBstMidEndSepPunct{\mcitedefaultmidpunct}
{\mcitedefaultendpunct}{\mcitedefaultseppunct}\relax
\EndOfBibitem
\bibitem[Legeza and S{\'{o}}lyom(2003)Legeza, and S{\'{o}}lyom]{Legeza2003}
Legeza,~O.; S{\'{o}}lyom,~J. Optimizing the density-matrix renormalization
  group method using quantum information entropy. \emph{Phys. Rev. B}
  \textbf{2003}, \emph{68}, 195116\relax
\mciteBstWouldAddEndPuncttrue
\mciteSetBstMidEndSepPunct{\mcitedefaultmidpunct}
{\mcitedefaultendpunct}{\mcitedefaultseppunct}\relax
\EndOfBibitem
\bibitem[Legeza and S{\'{o}}lyom(2004)Legeza, and S{\'{o}}lyom]{Legeza2004}
Legeza,~O.; S{\'{o}}lyom,~J. Quantum data compression, quantum information
  generation, and the density-matrix renormalization-group method. \emph{Phys.
  Rev. B} \textbf{2004}, \emph{70}, 205118\relax
\mciteBstWouldAddEndPuncttrue
\mciteSetBstMidEndSepPunct{\mcitedefaultmidpunct}
{\mcitedefaultendpunct}{\mcitedefaultseppunct}\relax
\EndOfBibitem
\bibitem[Bross \latin{et~al.}(2013)Bross, Hill, Werner, and
  Peterson]{Bross2013}
Bross,~D.~H.; Hill,~J.~G.; Werner,~H.-J.; Peterson,~K.~A. Explicitly correlated
  composite thermochemistry of transition metal species. \emph{J. Chem. Phys.}
  \textbf{2013}, \emph{139}, 094302\relax
\mciteBstWouldAddEndPuncttrue
\mciteSetBstMidEndSepPunct{\mcitedefaultmidpunct}
{\mcitedefaultendpunct}{\mcitedefaultseppunct}\relax
\EndOfBibitem
\bibitem[Ivanic \latin{et~al.}(2004)Ivanic, Collins, and Burt]{Ivanic2004}
Ivanic,~J.; Collins,~J.~R.; Burt,~S.~K. Theoretical Study of the Low Lying
  Electronic States of {oxoX}(salen) (X = Mn, Mn-, Fe, and Cr-) Complexes.
  \emph{J. Phys. Chem. A} \textbf{2004}, \emph{108}, 2314--2323\relax
\mciteBstWouldAddEndPuncttrue
\mciteSetBstMidEndSepPunct{\mcitedefaultmidpunct}
{\mcitedefaultendpunct}{\mcitedefaultseppunct}\relax
\EndOfBibitem
\bibitem[Ghosh \latin{et~al.}(2008)Ghosh, Hachmann, Yanai, and Chan]{Ghosh2008}
Ghosh,~D.; Hachmann,~J.; Yanai,~T.; Chan,~G. K.-L. Orbital optimization in the
  density matrix renormalization group, with applications to polyenes and
  $\beta$-carotene. \emph{J. Chem. Phys.} \textbf{2008}, \emph{128},
  144117\relax
\mciteBstWouldAddEndPuncttrue
\mciteSetBstMidEndSepPunct{\mcitedefaultmidpunct}
{\mcitedefaultendpunct}{\mcitedefaultseppunct}\relax
\EndOfBibitem
\bibitem[Zgid and Nooijen(2008)Zgid, and Nooijen]{Zgid2008}
Zgid,~D.; Nooijen,~M. The density matrix renormalization group self-consistent
  field method: Orbital optimization with the density matrix renormalization
  group method in the active space. \emph{J. Chem. Phys.} \textbf{2008},
  \emph{128}, 144116\relax
\mciteBstWouldAddEndPuncttrue
\mciteSetBstMidEndSepPunct{\mcitedefaultmidpunct}
{\mcitedefaultendpunct}{\mcitedefaultseppunct}\relax
\EndOfBibitem
\bibitem[Yanai \latin{et~al.}(2009)Yanai, Kurashige, Ghosh, and
  Chan]{Yanai2009}
Yanai,~T.; Kurashige,~Y.; Ghosh,~D.; Chan,~G. K.-L. Accelerating convergence in
  iterative solution for large-scale complete active space
  self-consistent-field calculations. \emph{Int. J. Quantum Chem.}
  \textbf{2009}, \emph{109}, 2178--2190\relax
\mciteBstWouldAddEndPuncttrue
\mciteSetBstMidEndSepPunct{\mcitedefaultmidpunct}
{\mcitedefaultendpunct}{\mcitedefaultseppunct}\relax
\EndOfBibitem
\bibitem[Dunning(1989)]{Dunning1989}
Dunning,~T.~H. Gaussian basis sets for use in correlated molecular
  calculations. I. The atoms boron through neon and hydrogen. \emph{J. Chem.
  Phys.} \textbf{1989}, \emph{90}, 1007--1023\relax
\mciteBstWouldAddEndPuncttrue
\mciteSetBstMidEndSepPunct{\mcitedefaultmidpunct}
{\mcitedefaultendpunct}{\mcitedefaultseppunct}\relax
\EndOfBibitem
\bibitem[Woon and Dunning(1993)Woon, and Dunning]{Woon1993}
Woon,~D.~E.; Dunning,~T.~H. Gaussian basis sets for use in correlated molecular
  calculations. {III}. The atoms aluminum through argon. \emph{J. Chem. Phys.}
  \textbf{1993}, \emph{98}, 1358--1371\relax
\mciteBstWouldAddEndPuncttrue
\mciteSetBstMidEndSepPunct{\mcitedefaultmidpunct}
{\mcitedefaultendpunct}{\mcitedefaultseppunct}\relax
\EndOfBibitem
\bibitem[Balabanov and Peterson(2005)Balabanov, and Peterson]{Balabanov2005}
Balabanov,~N.~B.; Peterson,~K.~A. Systematically convergent basis sets for
  transition metals. I. All-electron correlation consistent basis sets for the
  3d elements Sc{\textendash}Zn. \emph{J. Chem. Phys.} \textbf{2005},
  \emph{123}, 064107\relax
\mciteBstWouldAddEndPuncttrue
\mciteSetBstMidEndSepPunct{\mcitedefaultmidpunct}
{\mcitedefaultendpunct}{\mcitedefaultseppunct}\relax
\EndOfBibitem
\bibitem[Weigend \latin{et~al.}(2002)Weigend, K\"{o}hn, and
  H\"{a}ttig]{Weigend2002}
Weigend,~F.; K\"{o}hn,~A.; H\"{a}ttig,~C. Efficient use of the correlation
  consistent basis sets in resolution of the identity {MP}2 calculations.
  \emph{J. Chem. Phys.} \textbf{2002}, \emph{116}, 3175--3183\relax
\mciteBstWouldAddEndPuncttrue
\mciteSetBstMidEndSepPunct{\mcitedefaultmidpunct}
{\mcitedefaultendpunct}{\mcitedefaultseppunct}\relax
\EndOfBibitem
\bibitem[Manni and Alavi(2018)Manni, and Alavi]{LiManni2018}
Manni,~G.~L.; Alavi,~A. Understanding the Mechanism Stabilizing Intermediate
  Spin States in Fe({II})-Porphyrin. \emph{J. Phys. Chem. A} \textbf{2018},
  \emph{122}, 4935--4947\relax
\mciteBstWouldAddEndPuncttrue
\mciteSetBstMidEndSepPunct{\mcitedefaultmidpunct}
{\mcitedefaultendpunct}{\mcitedefaultseppunct}\relax
\EndOfBibitem
\bibitem[Weigend and Ahlrichs(2005)Weigend, and Ahlrichs]{Weigend2005}
Weigend,~F.; Ahlrichs,~R. Balanced basis sets of split valence, triple zeta
  valence and quadruple zeta valence quality for H to Rn: Design and assessment
  of accuracy. \emph{Phys. Chem. Chem. Phys.} \textbf{2005}, \emph{7},
  3297\relax
\mciteBstWouldAddEndPuncttrue
\mciteSetBstMidEndSepPunct{\mcitedefaultmidpunct}
{\mcitedefaultendpunct}{\mcitedefaultseppunct}\relax
\EndOfBibitem
\bibitem[Hellweg \latin{et~al.}(2007)Hellweg, H\"{a}ttig, H\"{o}fener, and
  Klopper]{Hellweg2007}
Hellweg,~A.; H\"{a}ttig,~C.; H\"{o}fener,~S.; Klopper,~W. Optimized accurate
  auxiliary basis sets for {RI}-{MP}2 and {RI}-{CC}2 calculations for the atoms
  Rb to Rn. \emph{Theor. Chem. Acc.} \textbf{2007}, \emph{117}, 587--597\relax
\mciteBstWouldAddEndPuncttrue
\mciteSetBstMidEndSepPunct{\mcitedefaultmidpunct}
{\mcitedefaultendpunct}{\mcitedefaultseppunct}\relax
\EndOfBibitem
\bibitem[Pittner \latin{et~al.}(2001)Pittner, Nachtigall, \v{C}\'{a}rsky, and
  Huba\v{c}]{Pittner2001}
Pittner,~J.; Nachtigall,~P.; \v{C}\'{a}rsky,~P.; Huba\v{c},~I. State-Specific
  Brillouin-Wigner Multireference Coupled Cluster Study of the Singlet-Triplet
  Separation in the Tetramethyleneethane Diradical. \emph{J. Phys. Chem. A}
  \textbf{2001}, \emph{105}, 1354--1356\relax
\mciteBstWouldAddEndPuncttrue
\mciteSetBstMidEndSepPunct{\mcitedefaultmidpunct}
{\mcitedefaultendpunct}{\mcitedefaultseppunct}\relax
\EndOfBibitem
\bibitem[Bhaskaran-Nair \latin{et~al.}(2011)Bhaskaran-Nair, Demel,
  {\v{S}}mydke, and Pittner]{BhaskaranNair2011}
Bhaskaran-Nair,~K.; Demel,~O.; {\v{S}}mydke,~J.; Pittner,~J. Multireference
  state-specific Mukherjee's coupled cluster method with noniterative
  triexcitations using uncoupled approximation. \emph{J. Chem. Phys.}
  \textbf{2011}, \emph{134}, 154106\relax
\mciteBstWouldAddEndPuncttrue
\mciteSetBstMidEndSepPunct{\mcitedefaultmidpunct}
{\mcitedefaultendpunct}{\mcitedefaultseppunct}\relax
\EndOfBibitem
\bibitem[Chattopadhyay \latin{et~al.}(2011)Chattopadhyay, Chaudhuri, and
  Mahapatra]{Chattopadhyay2011}
Chattopadhyay,~S.; Chaudhuri,~R.~K.; Mahapatra,~U.~S. Ab Initio Multireference
  Investigation of Disjoint Diradicals: Singlet versus Triplet Ground States.
  \emph{{ChemPhysChem}} \textbf{2011}, \emph{12}, 2791--2797\relax
\mciteBstWouldAddEndPuncttrue
\mciteSetBstMidEndSepPunct{\mcitedefaultmidpunct}
{\mcitedefaultendpunct}{\mcitedefaultseppunct}\relax
\EndOfBibitem
\bibitem[Pozun \latin{et~al.}(2013)Pozun, Su, and Jordan]{Pozun2013}
Pozun,~Z.~D.; Su,~X.; Jordan,~K.~D. Establishing the Ground State of the
  Disjoint Diradical Tetramethyleneethane with Quantum Monte Carlo. \emph{J.
  Am. Chem. Soc.} \textbf{2013}, \emph{135}, 13862--13869\relax
\mciteBstWouldAddEndPuncttrue
\mciteSetBstMidEndSepPunct{\mcitedefaultmidpunct}
{\mcitedefaultendpunct}{\mcitedefaultseppunct}\relax
\EndOfBibitem
\bibitem[Zhang \latin{et~al.}(1990)Zhang, Loebach, Wilson, and
  Jacobsen]{Zhang1990}
Zhang,~W.; Loebach,~J.~L.; Wilson,~S.~R.; Jacobsen,~E.~N. Enantioselective
  epoxidation of unfunctionalized olefins catalyzed by salen manganese
  complexes. \emph{J. Am. Chem. Soc.} \textbf{1990}, \emph{112},
  2801--2803\relax
\mciteBstWouldAddEndPuncttrue
\mciteSetBstMidEndSepPunct{\mcitedefaultmidpunct}
{\mcitedefaultendpunct}{\mcitedefaultseppunct}\relax
\EndOfBibitem
\bibitem[Irie \latin{et~al.}(1990)Irie, Noda, Ito, Matsumoto, and
  Katsuki]{Irie1990}
Irie,~R.; Noda,~K.; Ito,~Y.; Matsumoto,~N.; Katsuki,~T. Catalytic asymmetric
  epoxidation of unfunctionalized olefins. \emph{Tetrahedron Lett.}
  \textbf{1990}, \emph{31}, 7345--7348\relax
\mciteBstWouldAddEndPuncttrue
\mciteSetBstMidEndSepPunct{\mcitedefaultmidpunct}
{\mcitedefaultendpunct}{\mcitedefaultseppunct}\relax
\EndOfBibitem
\bibitem[Sears and Sherrill(2006)Sears, and Sherrill]{Sears2006}
Sears,~J.~S.; Sherrill,~C.~D. The electronic structure of oxo-Mn(salen):
  Single-reference and multireference approaches. \emph{J. Chem. Phys.}
  \textbf{2006}, \emph{124}, 144314\relax
\mciteBstWouldAddEndPuncttrue
\mciteSetBstMidEndSepPunct{\mcitedefaultmidpunct}
{\mcitedefaultendpunct}{\mcitedefaultseppunct}\relax
\EndOfBibitem
\bibitem[Ma \latin{et~al.}(2011)Ma, Manni, and Gagliardi]{Ma2011}
Ma,~D.; Manni,~G.~L.; Gagliardi,~L. The generalized active space concept in
  multiconfigurational self-consistent field methods. \emph{J. Chem. Phys.}
  \textbf{2011}, \emph{135}, 044128\relax
\mciteBstWouldAddEndPuncttrue
\mciteSetBstMidEndSepPunct{\mcitedefaultmidpunct}
{\mcitedefaultendpunct}{\mcitedefaultseppunct}\relax
\EndOfBibitem
\bibitem[Wouters \latin{et~al.}(2014)Wouters, Bogaerts, Voort, Speybroeck, and
  Neck]{Wouters2014_oxo}
Wouters,~S.; Bogaerts,~T.; Voort,~P. V.~D.; Speybroeck,~V.~V.; Neck,~D.~V.
  Communication: {DMRG}-{SCF} study of the singlet, triplet, and quintet states
  of oxo-Mn(Salen). \emph{J. Chem. Phys.} \textbf{2014}, \emph{140},
  241103\relax
\mciteBstWouldAddEndPuncttrue
\mciteSetBstMidEndSepPunct{\mcitedefaultmidpunct}
{\mcitedefaultendpunct}{\mcitedefaultseppunct}\relax
\EndOfBibitem
\bibitem[Stein and Reiher(2016)Stein, and Reiher]{Stein2016}
Stein,~C.~J.; Reiher,~M. Automated Selection of Active Orbital Spaces. \emph{J.
  Chem. Theory Comput.} \textbf{2016}, \emph{12}, 1760--1771\relax
\mciteBstWouldAddEndPuncttrue
\mciteSetBstMidEndSepPunct{\mcitedefaultmidpunct}
{\mcitedefaultendpunct}{\mcitedefaultseppunct}\relax
\EndOfBibitem
\bibitem[Sharma \latin{et~al.}(2017)Sharma, Knizia, Guo, and Alavi]{Sharma2017}
Sharma,~S.; Knizia,~G.; Guo,~S.; Alavi,~A. Combining Internally Contracted
  States and Matrix Product States To Perform Multireference Perturbation
  Theory. \emph{J. Chem. Theory Comput.} \textbf{2017}, \emph{13},
  488--498\relax
\mciteBstWouldAddEndPuncttrue
\mciteSetBstMidEndSepPunct{\mcitedefaultmidpunct}
{\mcitedefaultendpunct}{\mcitedefaultseppunct}\relax
\EndOfBibitem
\bibitem[Manni \latin{et~al.}(2019)Manni, Kats, Tew, and Alavi]{LiManni2019}
Manni,~G.~L.; Kats,~D.; Tew,~D.~P.; Alavi,~A. Role of Valence and Semicore
  Electron Correlation on Spin Gaps in Fe({II})-Porphyrins. \emph{J. Chem.
  Theory Comput.} \textbf{2019}, \emph{15}, 1492--1497\relax
\mciteBstWouldAddEndPuncttrue
\mciteSetBstMidEndSepPunct{\mcitedefaultmidpunct}
{\mcitedefaultendpunct}{\mcitedefaultseppunct}\relax
\EndOfBibitem
\end{mcitethebibliography}
\bibliographystyle{achemso}

\end{document}